\documentclass[aps, prd, 10pt, a4paper, twocolumn, nofootinbib, showpacs, amsfonts]{revtex4-2}

\usepackage[utf8]{inputenc}
\usepackage{dsfont}
\usepackage{dcolumn}
\usepackage{amssymb}
\usepackage{graphicx}
\usepackage{amsmath}
\usepackage{bm}
\usepackage{xcolor}
\usepackage{mathtools}
\usepackage{tabularx}
\usepackage{lipsum}
\usepackage{amsmath}
\usepackage{enumitem}
\usepackage{hyperref}
\usepackage{booktabs}
\usepackage[font=small]{caption}
\usepackage{tikz}
\usepackage{empheq,etoolbox}
\patchcmd{\subequations}
  {\theparentequation\alph{equation}}
  {\theparentequation.\alph{equation}}
  {}{}

\def\checkmark{\tikz\fill[scale=0.4](0,.35) -- (.25,0) -- (1,.7) -- (.25,.15) -- cycle;}
\newcommand\crossmark[1][]{%
  \tikz[scale=0.4,#1]{
    \fill(0,0)--(0.1,0) .. controls (0.5,0.4) .. (1,0.7)--(0.9,0.7) ..  controls (0.5,0.5) ..(0,0.1) --cycle;
    \fill(1,0.1)--(0.9,0.1) .. controls (0.5,0.3) .. (0,0.7)--(0.1,0.7) .. controls (0.5,0.4) ..(1,0.2) --cycle;
  }%
}
\usepackage{tikz}
\usepackage{mathrsfs}
\usetikzlibrary{arrows}

\usepackage{cases}
\patchcmd\subequations
 {\theparentequation\alph{equation}}
 {\subequationsformat}
 {}{}

\usetikzlibrary{calc,decorations.pathmorphing,shapes}

\usepackage{fancyhdr}
\newcommand{\bea}{\begin{eqnarray}}
\newcommand{\eea}{\end{eqnarray}}
\newcommand{\be}{\begin{equation}}
\newcommand{\ee}{\end{equation}}

\definecolor{rvwvcq}{rgb}{0.08235294117647059,0.396078431372549,0.7529411764705882}
\definecolor{wrwrwr}{rgb}{0.3803921568627451,0.3803921568627451,0.3803921568627451}

\begin{document}

\preprint{APS/123-QED}

\title{Regular Black Hole Cores\\
via Gravitational Evanescence of Collapsing Matter}

\author{Antonio Panassiti}
\email{antonio.panassiti@phd.unict.it}
\affiliation{INAF, Osservatorio Astrofisico di Catania, via S.Sofia 78, I-95123 Catania, Italy}
\affiliation{Dipartimento di Fisica e Astronomia “Ettore Majorana”, Universit\`a di Catania, Via S.
Sofia 64, 95123, Catania, Italy}
\affiliation{High Energy Physics Department, Institute of Mathematics, Astrophysics, and Particle Physics, Radboud University, Heyendalseweg 135, 6525 AJ Nijmegen, The Netherlands}

\date{\today}

\begin{abstract}
Modified Einstein's equations implementing both an energy-dependent Newtonian and cosmological constant can be obtained via a modified action characterised by a non-minimal gravity-matter coupling. We show how different dynamics for the high energy regime - that nevertheless have in common the vanishing of the Newton constant - result in models of gravitational collapse which source the formation of non-singular geometries with different types of asymptotic core: de Sitter, Minkowski, and with “steep pressure”. As we categorise their defining properties, one feature, which we prove to be actually independent of both the specific form of the non-minimal coupling and the equation of state of the collapsing fluid, stands out: the formation of a geometry with Minkowski core occurs only if the Newton constant, before finally vanishing, assumes negative values.
\end{abstract}

\maketitle

\section{Introduction and Motivation}
To date, our most successful theory for the physics of the gravitational interaction is General Relativity (GR) \cite{Will:2014kxa}. The formation of black holes (BHs) due to complete gravitational collapse, originally proposed in the Oppenheimer-Snyder-Datt (OSD) model \cite{Oppenheimer:1939ue, 1938ZPhy..108..314D}, is one of the most fascinating predictions of the theory. In the last decades, the research regarding these objects has rapidly expanded from the theoretical to the observational realm \cite{Buoninfante:2024oxl, Berti:2025hly, LIGOScientific:2016aoc, LIGOScientific:2016sjg, LIGOScientific:2017bnn, EventHorizonTelescope:2019dse, EventHorizonTelescope:2022wkp, EventHorizonTelescope:2022xqj}. However, BHs come with a shortcoming for the theory itself: the presence of curvature singularities, whose teleological occurrence is established by the singularity theorem \cite{Hawking:1970zqf}.

It is commonly believed that a fundamental theory of quantum gravity and matter should lead to a more refined description of gravitational collapse, and to the avoidance of singularity formation \cite{Bojowald:2007ky, Bosma:2019aiu, Kuntz:2019lzq, Buoninfante:2024yth}. While such a complete theory is not yet available as established science \cite{Kiefer:2025udf, Bambi:2023jiz, Loll:2022ibq}, an effective approach to the singularity problem is the introduction of regularized versions of classical BHs, having a finite value of the curvature at all points in spacetime \cite{Ansoldi:2008jw, Lan:2023cvz, Bambi:2023try, Carballo-Rubio:2025fnc}. These non-singular classical geometries, which can represent either regular BHs or horizonless BH mimickers \cite{Carballo-Rubio:2022nuj, Casadio:2024lgw}, constitute a heuristic tool to explore possible theoretical and phenomenological deviations from GR in the strong-gravity regime \cite{Ayon-Beato:1999kuh, Bojowald:2005qw, Hossenfelder:2009fc, Casadio:2010fw, Bambi:2013caa, Bambi:2013gva,Liu:2014kra, Torres:2014gta, Saueressig:2015xua, Adeifeoba:2018ydh, Carballo-Rubio:2018jzw, Carballo-Rubio:2019fnb, Platania:2019kyx, Husain:2021ojz, Li:2022eue, Lewandowski:2022zce, Malafarina:2022oka, Franzin:2022iai, Vellucci:2022hpl, Sau:2022afl, Kamenshchik:2023woo, Cadoni:2023lum, Cadoni:2023lqe, Franzin:2023slm, Boos:2023xoq, Cipriani:2024nhx, Carballo-Rubio:2024rlr, Arrechea:2024nlp, Delaporte:2024and, Sanchez:2024nub, Sanchez:2024sdm, Stashko:2024wuq, Konoplya:2024lch, Coviello:2025pla, Bonanno:2025bgc}. Furthermore, they represent a chance to experimentally test such deviations \cite{Bambi:2015kza, Yagi:2016jml, Bambi:2017khi, Riaz:2022rlx, Maselli:2023khq, Carballo-Rubio:2023ekp, Foschini:2024tgb, Sau:2025aqw}.
 
In order to obtain regular solutions, an old route, which is vastly explored in the literature, is that of modifying \textit{ad hoc} the static Misner-Sharp-Hernandez mass $M$ \cite{Misner:1964je, Hernandez:1966zia} in such a way that its behaviour in the asymptotic inner core reproduces that of a de Sitter space \cite{Markov:1982, Markov:1984ii, Frolov:1989pf, Frolov:1988vj, Frolov:2016pav}, as $M(R)=O(R^3)$, with $R$ being the radial Schwarzschild coordinate in a spherically symmetric setting. Such a rapid convergence to zero ensures the curvature to remain finite at $R=0$. For instance, the Bardeen metric \cite{1968qtr..conf...87B}, the Dymnikova spacetime \cite{Dymnikova:1992ux, Dymnikova:2001fb}, the Bonanno-Reuter BH \cite{Bonanno:2000ep}, and the Hayward geometry \cite{Hayward:2005gi}, they all fall in this category (see also \cite{Ayon-Beato:1999qin, Fan:2016hvf, Cadoni:2022chn} for other examples).

A different way to avoid the curvature divergence is pursued by the Ghosh-Culetu \cite{Ghosh:2014pba, Culetu:2014lca} and Simpson-Visser BHs \cite{Simpson:2019mud}, having a less traditional Minkowski core. In this case $M(R)=o(R^{3})$, \textit{i.e.}\ the mass function expansion, in the small $R$ regime, starts from a term of order strictly higher than three.

More recently, in the context of a scenario inspired by the Asymptotic Safety program for quantum gravity \cite{Percacci:2017fkn, Reuter:2019byg, Bonanno:2020bil}, it has been proposed a regular BH whose $M(R)$ vanishes slower than the two aforementioned cases, due to the leading order of its expansion involving a logarithm \cite{Bonanno:2023rzk}. In the latter case, to fully understand how the curvature divergence is avoided, it can not be ignored that the geometry is generated by a collapsing distribution of matter - a ball of perfect fluid - which remains of finite size for all physically relevant observers.

Thus, the main aim of this work is to define and compare the physical features distinguishing different core types, and to trace where the physical mechanism of their formation lies. Compared to previous cases, we will show that, taking into account the dynamical process of gravitational collapse, a third possibility, which we denote as “\textit{steep pressure}” core, can arise. Moreover, we shall show how the origin of the differences is naturally linked to the type of dynamics governing the non-static process of collapse, and to the manner in which the validity of the strong energy condition (SEC) \cite{Curiel:2014zba}, a hypothesis of the singularity theorem, is violated.

In general, to derive regular BHs within a fundamental framework has proven to be a challenge \cite{Borissova:2020knn, Giacchini:2021pmr, Knorr:2022kqp, Jia:2022gzo, Borissova:2023kzq, Giacchini:2024exc}. For instance, on the one hand, those obtained from the Einstein-Hilbert action with non-linear electrodynamics as source \cite{Dymnikova:2004zc}, ultimately imply an undesirable multi-valued Lagrangian (see figure 1 in the same reference). On the other hand, when regular geometries are obtained considering the process of collapse but using equations which are not directly related to a Lagrangian, issues like the non-conservation of the total energy-momentum tensor can arise. Considering an example, this occurs in the effective model proposed in \cite{Bambi:2013caa}, where a cutoff in the energy-density is implemented at the level of the field equations only. Moreover, other models of collapse, which have been recently proposed in the context of quasi-topological gravity \cite{Bueno:2024eig} and Lovelock theory \cite{Fernandes:2025eoc, Fernandes:2025fnz}, stem from a variational principle as well, but either they do not rely upon four spacetime dimensions and a finite number of free parameters, or their solutions are not asymptotically flat. The models of non-singular gravitational collapse presented in this work are still based on the general method introduced in \cite{Bonanno:2023rzk}, which notably has an action as starting point, and through our study, we manage to avoid the aforementioned problems.

The action we adopt here represents a modified theory of gravity which introduces a non-minimal coupling $\chi$ between matter and geometry. The latter is a scalar function which depends on the matter energy-density. It induces modified field equations with varying Newton and cosmological constant, which we denote as $G$ and $\Lambda$, to be distinguished from $G_{\text{N}}$ and $\Lambda_0$, even though they have the same dimensionality. For low energy-density, we assume $\chi \approx 8\pi G_{\text{N}}$, as necessary to reproduce the first stages of the OSD model. For the high energy regime, we require $\chi \to 0$, and $G \to 0$, since we want to study the consequences of a vanishing Newton coupling as regularization condition. In the past literature, this approach has been linked to the Asymptotic Safety scenario, either for conceptual considerations only or to look for guidance in fixing the intermediate behaviour of the function $\chi$ \cite{Bonanno:2017gji, Bonanno:2019ilz, Platania:2020lqb, Bonanno:2023rzk}. However, the antiscreening behaviour of gravity at high energy could be a more general quantum gravitational feature \cite{Polyakov:1993tp}. At the same time, a classical but non-minimal gravity-matter interaction is on its own a possibility worth to explore \cite{CANTATA:2021asi}. Throughout this work we still espouse the idea of asymptotic freedom for gravity-matter systems as fundamental physical cause of the weakening of $G$, but we focus on the consequences of having different modalities in which such a weakening can occur, finding that they are intertwined with the different defining features of the non-singular cores.

As starting point of our study, we take full advantage of the feature encoded in the modified action, that other covariant proposals are not suitable to encode: the physical Newton and cosmological couplings run with respect to the physical energy-density of the collapsing matter fluid. Then, at the heart, we present, in complementarity with the defining features of the de Sitter and steep pressure core, a theorem which establishes that the formation of a spacetime with Minkowski core implies the occurrence of negative values of $G$ during the collapse. Finally, in order to exemplify the concepts introduced, as a case study we choose to show which one is the dynamics leading to Dymnikova-type black holes with de Sitter core. Indeed, when a dust equation of state for matter is assumed, our approach can encompass a large class of non-singular geometries, providing the possibility to identify the corresponding generating dynamics. Moreover, Dymnikova BHs have already been found in another framework employing a scale dependent Newton coupling \cite{Platania:2019kyx}. Through a generalization of this suitable case study, we discuss new geometries with Minkowski and steep pressure core.

The paper is organized in the following way. In Sect.\ \ref{Action} we introduce the modified action and field equations; in Sect.\ \ref{Interior} and Sect.\ \ref{Exterior} we present, respectively, the solution for the matter interior and for the exterior ending state, over a yet unspecified non-singular dynamics. The exterior geometry is obtained through the junction conditions at the matter ball's boundary surface. In Sect.\ \ref{Nonsingularcores} we discuss the defining properties of each core in relation to dynamics, proving the aforementioned general feature of the Minkowski case, and we present a case study for each type of core. Sect.\ \ref{Conclusion} is devoted to a critical assessment of our findings and to comments on future perspectives. To make the paper self-contained, details on the junction procedure are given in Appendix \ref{Appendix1}. We work in the system of units $c=1$, and with metric signature $[-, +,+,+]$.

\section{Markov-Mukhanov Action}\label{Action}
Following the idea introduced in \cite{Markov:1985py} by Markov and Mukhanov, the asymptotical disappearance of gravity at high energy for an hydrodynamical system, can be implemented consistently starting from the modified theory
\be\label{theory}
S = \frac{1}{16 \pi G_{\text{N}}} \int d^4 x \sqrt{-g} \left[\mathcal{R} + 2  \chi(\epsilon)  \mathcal{L}^{(m)} \right] \, \, ,
\ee
where $\epsilon$ is defined as the proper energy-density of the infinitesimal unit element of a matter fluid, characterized by its four velocity $u^\mu$ in a general reference frame $\{x^{\mu}\}$, such that $u_\mu u^\mu = -1$, while its rest-mass density is denoted as $\rho$. $\mathcal{L}^{(m)} = -\epsilon$ is the matter Lagrangian \cite{Markov:1985py, Minazzoli:2012md}, and $\chi = \chi(\epsilon)$ is a scalar function representing a non-minimal gravity-matter coupling. As proven in \cite{Harko:2010zi, Minazzoli:2012md}, the form of the Lagrangian of a barotropic perfect fluid is independent of both the relativistic modified theory with non-minimal coupling at hand and the equation of state of the fluid. This holds as long as (a) $\mathcal{L}^{(m)}$ does not depend on the derivatives of the metric, and (b) particle number of the fluid is conserved, $\nabla_\mu (\rho u^\mu)=0$, according to basic thermodynamic assumptions. At this level, the explicit form of $\chi(\epsilon)$ is not yet specified, but, to ensure consistency with the OSD model within GR, we assume $\chi(\epsilon \to 0) \approx 8 \pi G_{\text{N}}$.

In order to make fully explicit the variation of the matter part of the action
\be \label{matterpartvariation}
\frac{1}{\sqrt{-g}}\,\delta
\left(2\,\sqrt{-g}\,\chi\,\epsilon \right) =2\frac{\partial (\chi\epsilon)}{{\partial\epsilon}}\delta\epsilon -\chi\,\epsilon\,g_{\mu\nu}\,\delta g^{\mu\nu} \, \, ,
\ee
some care is required \cite{Groen:2007zz, Markov:1985py, Harko:2010zi}. Assuming the mass-continuity equation $\nabla_\mu (\rho u^\mu)=0$, and taking into account that
\be
\delta \rho=(\rho/2)(g_{\mu\nu}+u_{\mu}u_{\nu})\delta g^{\mu \nu} \, \, ,
\ee
as well as the relative variations of density
\be\label{relative}
\delta\rho/\rho=\delta\epsilon/[p(\epsilon)+\epsilon] \, \, ,
\ee
where $p(\epsilon)$ is the fluid pressure, the variation of the proper energy-density can be expressed with respect to the variation of the metric
\be
\delta \epsilon=[\epsilon+p(\epsilon)](g_{\mu \nu}+u_{\mu}u_{\nu})\delta g^{\mu \nu} \, \, .
\ee
The latter has to be plugged into the right-hand side of \eqref{matterpartvariation}. As a result, the total variation of the action leads to the following field equations
\begin{equation}
\label{effectiveEQ}
R_{\mu\nu}- \frac{1}{2}g_{\mu\nu}R = \frac{\partial (\chi \epsilon)}{\partial\epsilon}
T_{\mu\nu}^{(m)}+\frac{\partial \chi}{\partial \epsilon} \epsilon^2 g_{\mu\nu} \equiv T_{\mu\nu}^{(\text{eff})} \, \, ,
\end{equation}
where the matter energy-momentum tensor reads, as usual,
\be
T_{\mu \nu}^{(m)}=[\epsilon+p(\epsilon)](u_{\mu}u_{\nu})+p(\epsilon)g_{\mu \nu} \, \, .
\ee
It is intuitive to identify
\begin{equation}
G(\epsilon)\equiv \frac{1}{8\pi}\frac{\partial (\chi \epsilon)}{\partial \epsilon}, \quad \Lambda(\epsilon)\equiv-\frac{\partial \chi}{\partial \epsilon} \epsilon^2,
\end{equation}
as, respectively, varying Newton and cosmological constant.

Here a couple of remarks are in order. First, the non-minimal coupling $\chi$ induces the variation of $G$ and $\Lambda$, and specifying one among the three functions amounts to determining the other two. Second, from the Bianchi identity, we have $\nabla ^{\mu} T_{\mu \nu}^{(\text{eff})}=0$, that reads
\be\label{effectivecontinuity}
\begin{aligned}
\frac{\partial(\epsilon \chi)}{\partial \epsilon}\nabla_{\mu}T^{\mu \nu \, (m)} +\frac{\partial^2(\epsilon \chi)}{\partial \epsilon^2}(\epsilon+p)\Phi^{\nu}=0 \, \, ,
\end{aligned}
\ee
with $\Phi^{\nu} \equiv (u^{\mu} u^{\nu}+g^{\mu \nu})\partial_{\mu}\epsilon$ \cite{Bonanno:2019ilz}. In the following, we shall assume that matter satisfies the usual conservation equation, $\nabla^{\mu}T_{\mu \nu}^{(m)}=0$, and to work with a homogeneous fluid. The latter  will source a Friedmann-Lemaître-Robertson-Walker (FLRW) spacetime, where in turn is ensured $\Phi^{\nu}=0$. These assumptions guarantee, for every choice of $\chi$, self-consistency with $\nabla ^{\mu} T_{\mu \nu}^{(\text{eff})}=0$, and, at the same time, compatibility with standard conservation of matter energy-momentum.\footnote{Thus, despite the non-minimal gravity-matter coupling, under these assumptions there is absence of “fifth force” and massive free falling particles are geodesics of the metric. This is not the standard situation in modified theories with non-minimal coupling (see, \textit{e.g.,}\ \cite{Sotiriou:2007zu} and references therein).}

The mechanism sourcing the running of the gravitational couplings in our framework, is not, in general, trivially related to the Asymptotic Safety program for quantum gravity. In that case, the scale dependence of dimensionless couplings is derived by applying exact functional renormalization group techniques to Einstein gravity. The non-perturbative setting allows to discover a non-trivial ultraviolet fixed point \cite{Reuter:1996cp}, which prevents the occurrence of divergences. However, the flow in theories-space is traced in terms of the euclidean momentum $k$. Eventually, in order to obtain the actual scale dependence of the physical couplings $G$ and $\Lambda$, in terms of the physical quantity which dictates the running (as $\epsilon$ in our framework), additional considerations are required. These additional steps constitute the “cutoff identification” issue \cite{Bonanno:2020bil}. While different rationales are available in the literature \cite{Falls:2010he, Koch:2013owa, Saueressig:2015xua, Platania:2019kyx, Borissova:2022mgd, Bonanno:2024wvb, Platania:2025imw}, in the collapse model in \cite{Bonanno:2023rzk}, $k$ was related to a distance scale $d$ interpreted as the proper distance in a spherically symmetric setting, as in \cite{Bonanno:2000ep}: $k \sim 1/d \sim \sqrt \epsilon$. When applied to the expression for $G(k)$ obtained from the Reuter fixed point, this identification provides one possible explicit map $G(k) \mapsto G(\epsilon)$, preserving the feature $G \to 0$ as $k \to \infty$ ($\epsilon \to \infty$). Instead, in this work, we shall select the specific functional form of the dynamics $\chi(\epsilon)$ according to a less reductionist rationale. Our procedure embeds in the collapse the property of asymptotic freedom for the gravity-matter system, but not necessarily a flow in theories-space as depicted by possible ultraviolet completion of gravity within the framework of a non-perturbative quantum field theory.

In this regard, the approach we have just introduced was originally employed in the context of cosmology by Markov, to encode into the dynamics the principle of a limiting density of matter as a universal law of nature, which in geometrical terms translates into the limiting curvature hypothesis \cite{Markov:1982}. Markov, based on the works of Fradkin and Vilkosky on the conformal invariant one-loop approximation of pure gravity \cite{Fradkin:1978, Fradkin:1978yf}, also pioneered the discussion on the relation of the model with the concept of asymptotic freedom for gravity \cite{Markov:1983, Markov:1984ii}. In particular, a limiting energy-density can be thought as the product of a fundamental energy dependence of the dynamics. At the level of an effective description, the right-hand side of Einstein equations could be altered in the strong-fields regime, either by semiclassical refinements of the vacuum expectation value, or by higher order curvature corrections. Similarities with the Born-Infeld formalism are also observed, since the limiting density principle would “propagate” from spacetime to all the fields living on it. Thus justified, the energy-dependent correction in the Einstein equations leads dynamically to a non-singular de Sitter-like initial state of the universe \cite{Markov:1982, Markov:1983, Markov:1984ii, Aman:1984}. Finally, Markov and Mukhanov proposed to derive the modification of the equations from the action in eq.\ \eqref{theory} \cite{Markov:1985py}. This last step essentially introduces the bound between the model and a non-minimal gravity-matter coupling, rooting the dependence on the energy-density at the deeper level of a theory.

\section{Interior Solution}\label{Interior}
In general, in a spherically symmetric context, the portion of spacetime permeated by a collapsing star, can be modelled using a metric ansatz in comoving coordinates $x^\mu=\{t, r, \theta, \phi\}$
\be\label{comovingansatz}
ds^2=-e^{-2 \nu (t, r)}dt^2+e^{2 \psi (t, r)}dr^2+C^{2} (t,r) d\Omega^2 \, \, ,
\ee
where $d\Omega^2$ represents the metric on the unit 2-sphere, sourced by a ball of perfect fluid, with energy-density $\epsilon=\epsilon(t, r)$ and pressure $p=p(t,r)$ \cite{Joshi:2008zz, Malafarina2016, Malafarina2023}. In our model, the aforementioned chart has a compact domain for the radial coordinate: $0 \leq r \leq r_{b}$, with $r_{b}$ being the location of the outermost shell layering the ball. Using \eqref{comovingansatz}, the independent field equations can be expressed as
\begin{align}
\label{G00}
\frac{F_{\rm eff}'}{C^2C'} &= 8 \pi G(\epsilon)\epsilon  + \Lambda(\epsilon) =\chi(\epsilon) \epsilon \equiv \epsilon_{\rm eff} \, \, , \\
\label{G11}
-\frac{\dot{F}_{\rm eff}}{C^2\dot{C}} &= 8 \pi G(\epsilon)p-\Lambda(\epsilon) \equiv p_{\rm eff} \, \, ,
\end{align}
and
\be
\label{G01}
\dot{C}'= \dot{C}\nu'+C'\dot{\psi} \, \, .
\ee
The function $F_{\text{eff}}(t,r)$ represents the effective Misner-Sharp-Hernandez mass, which has been defined, in analogy with the static case, requiring that $g_{\mu \nu} \nabla^{\mu}C \nabla^{\nu}C=1-F_{\text{eff}}/C$, which gives another independent equation
\be\label{effectiveMisner}
F_{\rm eff}=C(1-C'^2e^{-2\psi}+\dot{C}^2e^{-2\nu}) \, \, .
\ee
A dot represents a derivative with respect to the comoving time, while a prime denotes derivative with respect to the comoving radial coordinate. Finally, the Bianchi identity takes the form
\be\label{effectiveBianchi}
\nu'=-\frac{p_{\rm eff}'}{\epsilon_{\rm eff}+p_{\rm eff}} \, \, .
\ee

As mentioned before, we are interested in a homogeneous fluid, described by $\epsilon=\epsilon(t)$ and $p=p(t)$, which in turn imply $\epsilon_{\text{eff}}(t)$ and $p_{\text{eff}}=p_{\text{eff}}(t)$. As a consequence, from \eqref{effectiveBianchi}, we have $\nu=\nu(t)$, that can always be rescaled such that $\nu=0$. Then, integrating \eqref{G01}, we obtain $e^{2 \psi}=C'^2(1-Kr^2)$, where the integration constant $K$ represents the intrinsic curvature. The areal-radius function $C$ can always be rescaled as $C(t, r)=a(t)r$, where $a(t)$ is the dimensionless scale factor, due to a residual gauge freedom. Thus, having a homogeneous and isotropic matter source, the initial ansatz reduces to the FLRW metric
\be\label{FLRW}
ds^2=-dt^2+\frac{a^2(t)}{1-Kr^2}dr^2+a^2(t)r^2 d \Omega^2 \, \, ,
\ee
and the conservation equation for matter reads
\be\label{matterBianchi}
d\epsilon + 3[p(\epsilon)+\epsilon]d \ln{a}=0 \, \, .
\ee
Moreover, the requirement that at $t=t_{i}$, when the collapse starts, the geometry is sourced by a regular star configuration, with finite values of initial energy-density and pressure everywhere, and free from any trapped surface, induces the following scaling for the effective mass function
\be\label{effectiveMisner}
F_{\text{eff}}(t, r)=r^3m_{\text{eff}}(t) \, \, ,
\ee
with $m_{\text{eff}}(t=t_i)$ having a finite value. We can then rewrite \eqref{G00}, \eqref{G11} and \eqref{effectiveMisner} as
\begin{align}
& \label{G00effective} \epsilon_{\text{eff}}=\frac{3m_{\text{eff}}}{a^3} \, \, , \\
&p_{\text{eff}}=-\frac{\dot{m}_{\text{eff}}}{\dot{a}a^{2}} \, \, ,
\end{align}
and
\be\label{effectivemass}
m_{\text{eff}}=a(\dot{a}^2+K) \, \, .
\ee
Identifying the potential, $V\equiv-m_{\text{eff}}/a$, the latter one is readily rewritten as
\be\label{scalefactor}
\dot{a}=-\sqrt{-V-K} \, \, .
\ee
Once - specifying the form of $\chi(\epsilon)$ - the theory is fixed, the system is described by two differential equations and three unknowns. After an equation of state for matter is assigned, the only remaining unknown is the scale factor. Thus, as shown by \eqref{scalefactor}, the problem of finding the solution for a spacetime permeated by a collapsing ball of homogeneous fluid, is computationally equivalent to the problem of finding the “trajectory” $a(t)$ of a particle with initial energy $K$ and subjected to the potential $V(a)$.

In order to study properly the collapse, we need to introduce a few more ingredients. In a spherically symmetric spacetime, the formation of the apparent horizons, that characterizes the way in which the causal structure evolves, can be spotted by means of the condition $\nabla_{\mu}C \nabla^{\mu}C=0$ \cite{Ashtekar:2004cn, Helou:2016xyu, Faraoni:2016xgy, Ashtekar:2025wnu}, which using the definition of effective Misner-Sharp mass reads
\be
1-\frac{F_{\text{eff}}}{C}=0 \, \, .
\ee
Using \eqref{effectivemass}, the implicit solution of this equation is given by
\be\label{ahcurve}
r_{\text{ah}}(t)=\frac{1}{\sqrt{\dot{a}^2+K}} \, \, ,
\ee
which is a mathematical curve essentially describing at which comoving instant, the escape velocity from a certain comoving shell - layering the ball - equals the speed of light. The major indicator of the presence of a curvature singularity in the geometry is the Kretschmann scalar $\mathcal{K} \equiv \mathcal{R}_{\mu \nu \gamma \delta}\mathcal{R}^{\mu \nu \gamma \delta}$, which in the spacetime \eqref{FLRW} takes the form
\be\label{interiorKretschmann}
\mathcal{K}(a)=60\frac{m_{\text{eff}}^2}{a^6} \, \, .
\ee
The latter expression manifestly signals that, during the collapse, a singularity forms \textit{if the scale factor goes to zero in a finite amount of comoving time}. The discriminant for such an occurrence lies in the validity of the SEC. Indeed, if the SEC holds, as it happens in GR assigning a non-exotic equation of state for matter, $p(\epsilon) \geq 0$, all the hypotheses of the singularity theorem will be satisfied. Otherwise, the effective energy-momentum tensor, in the modified field equations, can violate the hypothesis regarding the validity of the SEC\footnote{The singularity theorem, in its most recent formulation \cite{Hawking:1970zqf}, assumes the SEC for the (effective) matter source.} even if matter obeys an ordinary equation of state. This opens up the possibility of avoiding the formation of the singularity. Thus, for our purposes, it is worth to write the SEC in terms of both the effective quantities and the varying constants,
\begin{subnumcases}{} \label{interiorSEC1}
\epsilon_{\text{eff}}=\chi(\epsilon)\epsilon \geq 0 \\ \label{interiorSEC2}
\epsilon_{\text{eff}}+p_{\text{eff}}=G(\epsilon)(\epsilon+p) \geq 0 \\ \label{interiorSEC3}
\epsilon_{\text{eff}}+3p_{\text{eff}}=G(\epsilon)(\epsilon+3p)-2\Lambda(\epsilon) \geq 0 \, \, .
\end{subnumcases}

At this stage, we clarify that gravitational evanescence of collapsing matter corresponds to a dynamics that fulfil the condition
\be\label{disappearance}
\lim_{\epsilon \to \infty} \chi(\epsilon)=0=\lim_{\epsilon \to \infty}G(\epsilon) \, \, .
\ee
Our approach allows to implement a regularization condition based on the scale-dependent behaviour of the gravity-matter system, thus, of the gravitational constant. The condition can also be expressed as
\be \label{regularpotential}
\lim_{a \to 0}V(a)= 0 = \lim_{\epsilon \to \infty}G(\epsilon) \, \, ,
\ee
indeed, the correlation between the non-minimal coupling and the potential is given by
\be\label{potential}
V(a)=- \frac{\epsilon(a)}{3} \chi(a) a^2 \, \, ,
\ee
and the increase of $\epsilon$ for decreasing values of $a$ is guaranteed by \eqref{matterBianchi} when a non-exotic equation of state for matter is assigned. Such a dynamics, which always violates the SEC, is non-singular if it also satisfies the aforementioned property for the scale factor. Finally, we remark that \eqref{disappearance} does not exclude that $G$, along its evolution, can be zero also at some finite energy-density value. In particular, this occurs if the condition $\Lambda(\epsilon)/\epsilon=\chi(\epsilon) \neq 0$, or alternatively $\epsilon_{\text{eff}}(\epsilon)=-p_{\text{eff}}(\epsilon) \neq 0$, is realized. While, in general, for $G \to 0$, the particles composing the ball of matter “become" less and less massive, in this case matter would not yet be definitely decoupled from the curvature, since $\chi$ would still be finite. Then, the moment $G$ changes its sign, the particles would start to “regain” their mass, but this time they would be repelled from each other.

\subsection*{Dust Matter and Parameters Space}
As anticipated, to close the system of differential equations we need to assign an equation of state.  Hereafter, except where otherwise specified, we work within the dust case, which amounts to neglecting interactions between the infinitesimal elements that compose the ball. From \eqref{relative} and \eqref{matterBianchi}, we have, respectively, $\rho=\epsilon$ and\footnote{A non-dust equation of state would introduce a difference between the rest and proper mass-density. Furthermore, similarly to the Lemaître-Tolman-Bondi collapse model in GR, it would make necessary to assume a non-homogeneous pressure, $p=p(r)$, to avoid a discontinuity at the boundary $r_b$ \cite{Joshi:2014gea}.}
\be \label{dust_energy_density}
\epsilon=\epsilon_{0}a^{-3} \, \, ,
\ee
where the integration constant $\epsilon_{0}$ is the energy-density at the initial time $t_i$, if $a(t_i)$ is set to unity.
\begin{figure*}[htbp!]
\centering
    \includegraphics[width=0.5\linewidth]{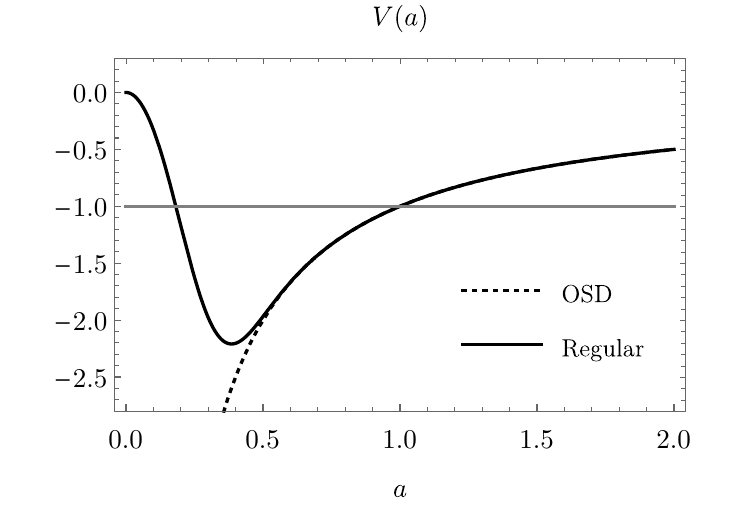}\hfil \hfil 
    \includegraphics[width=0.485\linewidth]{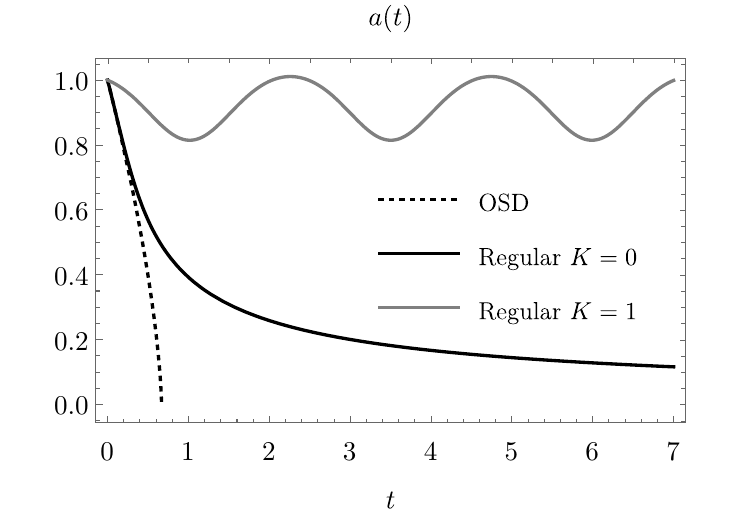}\par\medskip
\caption{Potential of the OSD collapse, for $m_{0}=1$, plotted against the potential of a non-singular dynamics; the intersections of the latter with the line $V(a)=-K$ occur at $a_{\text{min}}$ and $a_{\text{max}}$; for illustrative purpose we choose the dynamics \eqref{dynChip} with $p=2$, and $m_{0}=1$, $\xi=0.01$, however the qualitative behaviour holds true for all the non-singular cases (left panel). Scale factor for the OSD collapse, for $K=0$ and $m_{0}=1$, plotted against the bound, and marginally bound, non-singular collapse, for $p=3$, $m_{0}=1$ and $\xi=0.1$ (right panel). In all plots in the paper we set $8\pi G_{\text{N}}=1$.} \label{fig_interiorpotentialandscalefactor}
\end{figure*}

In order to implement condition \eqref{disappearance} and, at the same time, have consistency with GR in the low energy-density regime, it is necessary to introduce in $\chi(\epsilon)$ a new dimensionful scale $\xi$. This is the standard situation encountered whenever a possible regularization condition, that goes beyond GR, is proposed. We believe that the value of $\xi$ is related to a more fundamental theory, but, in our approach, it can be constrained a posteriori by observations only. This parameter can be thought as the inverse of the threshold energy at which the deviation from GR becomes relevant. Thus, it holds $\xi>0$, and we expect $\xi \ll 1/\epsilon_0$, and possibly, but not necessarily, $\xi \approx 1/ \epsilon_{\text{P}}$, with $\epsilon_{\text{P}}$ being the Planck density. The shape of the potential, which allows for singularity avoidance when a non-singular dynamics $\chi$ is specified, is displayed in the left panel of Fig.\ \ref{fig_interiorpotentialandscalefactor}. The initial condition $\epsilon_{0}$, as well as $\xi$, partly defines the parameters space of the theory. As we shall show in Sect.\ \ref{Exterior}, the latter plays an essential role in determining whether the geometry resulting from the collapse represents a regular BH or a horizonless BH mimicker. Further, the study of the parameters space will clarify that the mass parameter $M_0$, which appears in the static ending state, is actually not an independent one.
\begin{figure*}[htbp!]
\includegraphics[width=0.5\linewidth]{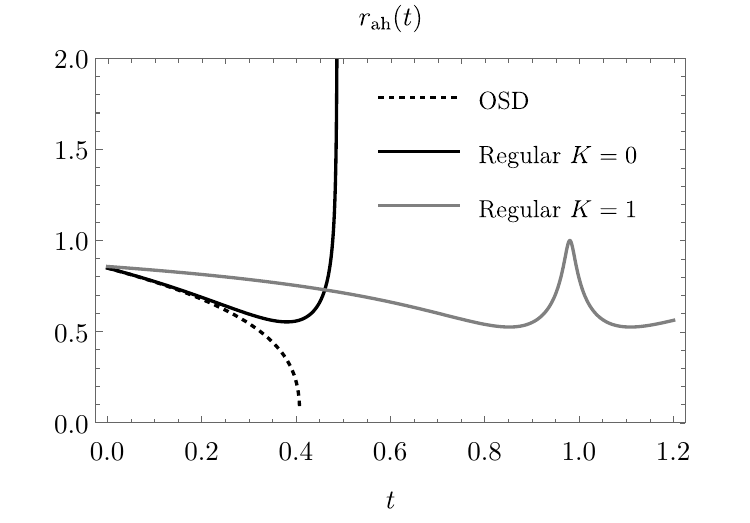}\hfil
\includegraphics[width=0.5\linewidth]{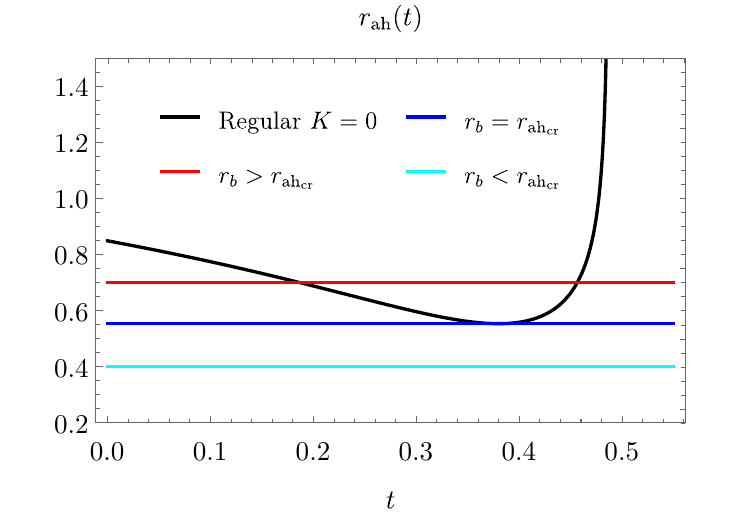}\par\medskip
\caption{Curve defined by \eqref{ahcurve} for the OSD model, with $m_0=1$, plotted against the one for the non-singular dynamics with $m_0=1$, $\xi=0.01$ and $p=3/2$ (left panel). The three different qualitative possibilities depending on the value of the parameter $r_{b}$. They correspond to different evolutions of the causal structure for the interior geometry (right panel). \label{fig_interiorapparenthorizon}}
\end{figure*}

At this point it is worth to remark that, usually, the limit - in parameters space - to GR of a modified gravity theory is a highly non-trivial issue \cite{Ovalle:2016pwp}. Ultimately, this is a subtle mathematical problem related to the high non-linearity of the field equations and to the change in the type of non-linearity occurring in a modified theory.\footnote{For instance, even a relatively simple modified theory as Brans-Dicke gravity \cite{Brans:1961sx, Brans:1962zz}, does not have a full limit to GR when the complete theory, instead of the weak-field model only, is considered \cite{Brans:2005ra, PanassitiMasterThesis}.} Remarkably, the Markov-Mukhanov approach circumvents such an issue, since does not introduce any new gravitational field beyond the metric, and since $\chi(\epsilon; \xi)$ does not alter, at any step, the original non-linearity of the Einstein field equations. The semicolon remarks, where needed, the dependence on parameters. Thus, if it holds, beyond $\chi(\epsilon \to 0; \xi) \approx 8\pi G_{\text{N}}$, also $\chi(\epsilon; \xi \to 0)=8\pi G_{\text{N}}$, then it is ensured at all levels, in the modified equations as well in their solutions, a full limit to GR in parameters space. In particular, in the limit $\xi \to 0$, \eqref{scalefactor} reduces to
\be
\dot{a}=-\sqrt{\frac{m_{0}}{a}-K} \, \, ,
\ee
the standard differential equation of the OSD model, whose potential and solution are displayed in Fig.\ \ref{fig_interiorpotentialandscalefactor}. We remark that, in general, the parameter $m_0$ that appears in the OSD collapse, constrained by the relation $3 m_{0}=8 \pi G_{\text{N}} \epsilon_{0}$, does not coincide with the value $m_{\text{eff}}(t=t_i)$. Then, for consistency, we have to require
\be \label{parameterlimit}
\lim_{\xi \to 0}m_{\text{eff}}(t; \xi)\equiv m_{0} \, \, ,
\ee
which has to hold independently of $t$. This choice is always possible due to the functional form of $m_{\text{eff}}$ with respect to $\chi$. Furthermore, the value $a(t=t_i)$ has an independent status as well. Indeed, although the scale factor is not an observable, the ratio $a(t)/{\dot{a}(t)}$, in principle, is. However, to simplify the study of the parameters space, we fix $a(t_i)=1$. Thus, currently, we are left with a three-dimensional space described by the coordinates $(\xi, r_b, \epsilon_0)$, or, alternatively, because of \eqref{parameterlimit}, by $(\xi, r_b, m_0)$. We recall that $r_b$ is the location of the boundary in comoving coordinate. Thus, it is related to the initial size of the matter ball. Finally, we can restrict the allowed values for $K$ to the discrete set $K=\{-1, 0, 1\}$, via an additional rescaling of the coordinates $t$ and $r$. In particular, $K=0$ means that, at $t=t_{i}$, the infalling particles composing the ball have zero velocity at radial infinity, $K=1$ that they have zero velocity at a finite radius, while $K=-1$ that they have non-zero velocity even at radial infinity. Hereafter, we exclude the negative value from our study, since we prefer to focus on the case of a matter source which is initially at rest.\footnote{Indeed, $K$ is related to the initial energy per unit mass of an infinitesimal element of matter fluid. This can be shown, for instance, using Lemaìtre coordinates \cite{Malafarina2023}.} The OSD collapse, regarding the ending state, does not qualitatively distinguish between the zero value and the positive one, meaning that it always produces a Schwarzschild BH. In our models with non-singular dynamics $\chi$, the matter distribution faces a different fate in the two cases.

\subsubsection*{Marginally Bound Collapse}
For $K=0$, the scale factor solution is of the type displayed in the right panel of Fig.\ \ref{fig_interiorpotentialandscalefactor}. Also, an analytical expression for $a(t)$ at large $t$ can be usually given, solving \eqref{scalefactor} with $V(a)$ replaced by its Taylor expansion around $a_{0}=0$. The main qualitative feature is that the scale factor decreases monotonously and never reaches zero in a finite amount of time. Since any physical observer can wait only for finite intervals of time, no observer will experience the singularity formation. Thus, in contrast to the OSD model, the star never degenerates into a geometrical point of zero volume. Even in the scenario in which the matter distribution reduces to a ball with radius of order of the Planck length, its volume is still non-zero, and its energy-density is still finite. What occurs is an \textit{eternal collapse}, with the strength of gravity progressively weakened as the contraction continues indefinitely. The time-reversal version of a similar solution is already known in cosmology as the “emergent universe” scenario \cite{Ellis:2002we, Zholdasbek:2024pxi}.

The corresponding curve \eqref{ahcurve}, which signals the formation of trapped surfaces, is shown in the left panel of Fig.\ \ref{fig_interiorapparenthorizon}. It reaches a minimum value $r_{{\text{ah}}_{\text{cr}}}=r_{{\text{ah}}_{\text{cr}}}(\xi, m_0)$, and then diverges. The minimum $r_{{\text{ah}}_{\text{cr}}}$ plays a discriminant role when it comes to depict the phase diagram structure within parameters space. Indeed, depending on the value of $r_b$, which must satisfy $r_b<r_{\text{ah}}(t_{i})$ (since we require no trapped surfaces when the collapse begins), there are three possible ways in which the causal structure of the interior solution can evolve (see the right panel of Fig.\ \ref{fig_interiorapparenthorizon}). If $r_b>r_{{\text{ah}}_{\text{cr}}}$, the boundary is solution of \eqref{ahcurve} twice. In particular, after the boundary becomes a trapped surface, some of the innermost shells layering the ball, up to the shell located at $r_{{\text{ah}}_{\text{cr}}}$, become trapped as well. Then, due to the potential $V(a)$ turning repulsive, these same shells return to being untrapped one by one, until the boundary itself returns to being untrapped. On the contrary, in the OSD collapse, all the innermost shells up to the center become trapped, and indeed the singularity forms when this occurs for the shell at $r=0$. If $r_b<r_{{\text{ah}}_{\text{cr}}}$, the spacetime is free of apparent horizons at all times. The extremal case $r_b=r_{{\text{ah}}_{\text{cr}}}$ means that only the boundary is solution of \eqref{ahcurve}, with multiplicity two with respect to $t$.
\begin{figure*}[htbp!]
\centering
    \includegraphics[width=0.41\linewidth]{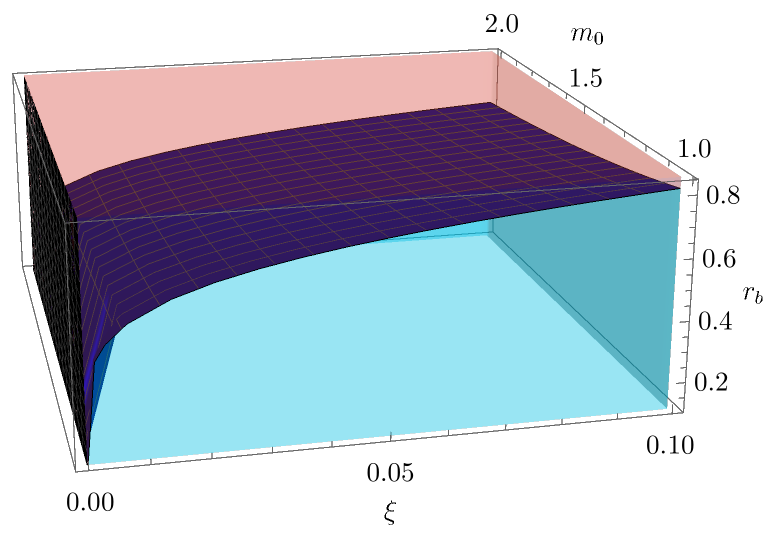} \hfill \hfill
    \includegraphics[width=0.16\linewidth]{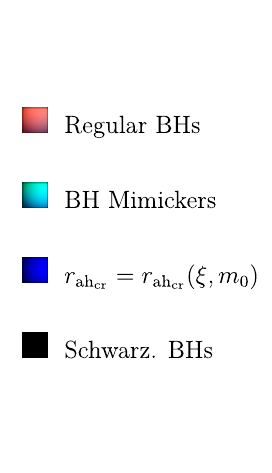}
    \includegraphics[width=0.41\linewidth]{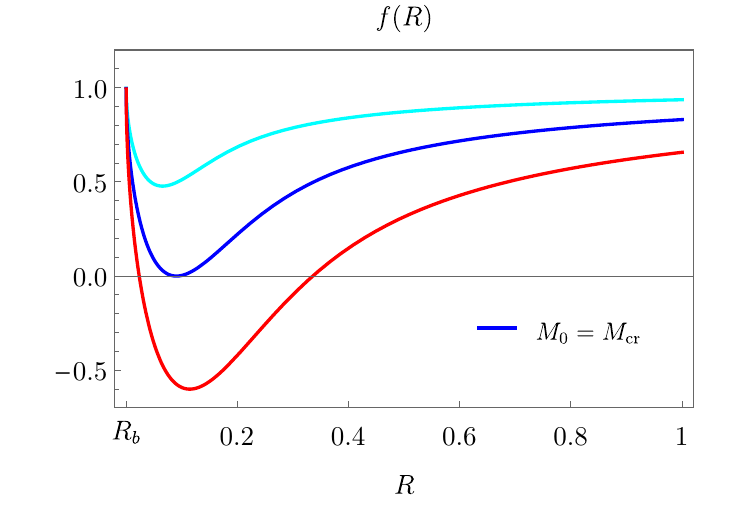}\par\medskip
\caption{Phase diagram of the parameters space of the model with dust equation of state and $K=0$, for the non-singular dynamics corresponding to $p=3/2$, however the qualitative structure holds true for all the non-singular cases (left panel). The three qualitatively different causal structures that are possible for the exterior geometry, for $m_0=1$ and $\xi=0.01$ ($r_{{\text{ah}}_{\text{cr}}}=0.553$, $M_{0_{\text{cr}}}=0.085$). The blue curve represents an extremal black hole (right panel). The points in parameters space and the possible causal structures of the global spacetime resulting from the collapse are related by a bijective map.} \label{fig_parameterscausal}
\end{figure*}

\subsubsection*{Bound Collapse}
For $K=1$, the matter distribution undergoes a \textit{bounce} at the instant $t_{\text{B}}$, for $a(t_{\text{B}})=a_{\text{min}}>0$, and a crunch at $2t_{\text{B}}$, for $a(2t_{\text{B}})=a_{\text{max}}$, as displayed in the right panel of Fig.\ \ref{fig_interiorpotentialandscalefactor}. Indeed, we have that the trajectory for $a(t)$ has two points of reversal of motion, where $\dot{a}=0$. In such an idealized model, the bounce and the crunch are symmetric, and the oscillation takes place endlessly with a period of $2t_{\text{B}}$. Clearly, the values of $t_{\text{B}}$, $a_{\text{min}}$, $a_{\text{max}}$ depend on $(\xi, r_b, m_0)$. The corresponding curve \eqref{ahcurve} is shown in the left panel of Fig.\ \ref{fig_interiorapparenthorizon}. Similarly to the previous case, depending on the value of $r_{b}$, three different qualitative evolutions of the causal structure are possible; indeed, it holds $r_{\text{ah}}(t_{\text{B}})>r_{\text{ah}}(t_{i})$ always. Finally, it is interesting to notice that effective models of collapse inspired by Loop Quantum Gravity, predict a bounce for both the marginally bound and bound case \cite{Han:2023wxg, Cafaro:2024lre}.

\section{Exterior Solution}\label{Exterior}
In order to determine the solution for the region of spacetime not permeated by the matter source, we employ the junction conditions at the boundary surface of the ball \cite{Lanczos:1922, Lanczos:1924, Darmois:1927, Israel:1966rt, Fayos:1992, Fayos:1996gw, Poisson:2009pwt}. In particular, for the exterior geometry we assume a generalized Schwarzschild ansatz in standard coordinates ${X^{\mu}}=\{ T, R, \theta, \phi\}$
\be\label{spacetimeexterior}
ds^2=-f(R)dT^2+f(R)^{-1}dR^2+R^2d\Omega^2 \, \, ,
\ee
where (hereafter, we use units $8\pi G_{\text{N}}=1$)
\be\label{lapseexterior}
f(R)\equiv1-2M(R)/R \, \, .
\ee
Here, the radial coordinate has domain $R \geq R_{b}$, where $R_{b}$ is the position of the boundary surface of the ball. The mass function $M(R)$, which is the unknown in our ansatz\footnote{In our model the Schwarzschild geometry is not allowed as solution for the exterior. This can be understood in many different ways: for instance, since in the interior it holds a varying mass function, $m_{\text{eff}}=m_{\text{eff}}(t)$, a constant mass $M_0$ in the exterior would imply an inconsistency in the effective flux of energy-momentum.}, has to be determined by the junction conditions themselves. From the point of view of the exterior geometry, the three-dimensional time-like boundary hypersurface $\Sigma$ is parametrized by the equation $R=R_{b}(T)$. Thus, the three-dimensional metric induced on $\Sigma$ is
\be
\begin{aligned}
ds^2\big|_{\Sigma} =
&-\left[f(R_b)-f(R_b)^{-1}\left(\frac{dR_b}{dT}\right)^2\right]dT^2 \\
&+R_b^2d\Omega^2 \, \, .
\end{aligned}
\ee
A similar computation of the induced metric from the point of view of the interior geometry, together with the matching conditions for the three-dimensional metric, give the implicit relation between the comoving time $t$ and the Schwarzschild time $T$ on $\Sigma$ (see Appendix \ref{Appendix1} for details), and the condition
\be\label{radialcoordinatematching}
R_{b}[T(t)]=a(t)r_{b} \, \, .
\ee
This establishes a further refinement for the domain of the radial coordinate of the exterior geometry:
\be\label{domainradial}
R \geq R_{b}>0 \, \, ,
\ee
where the last inequality follows from the result that $a(t)>0$ at all times, both for $K=0$ and $K=1$.
\begin{figure*}[htbp!]
\centering
    \includegraphics[width=0.26\linewidth]{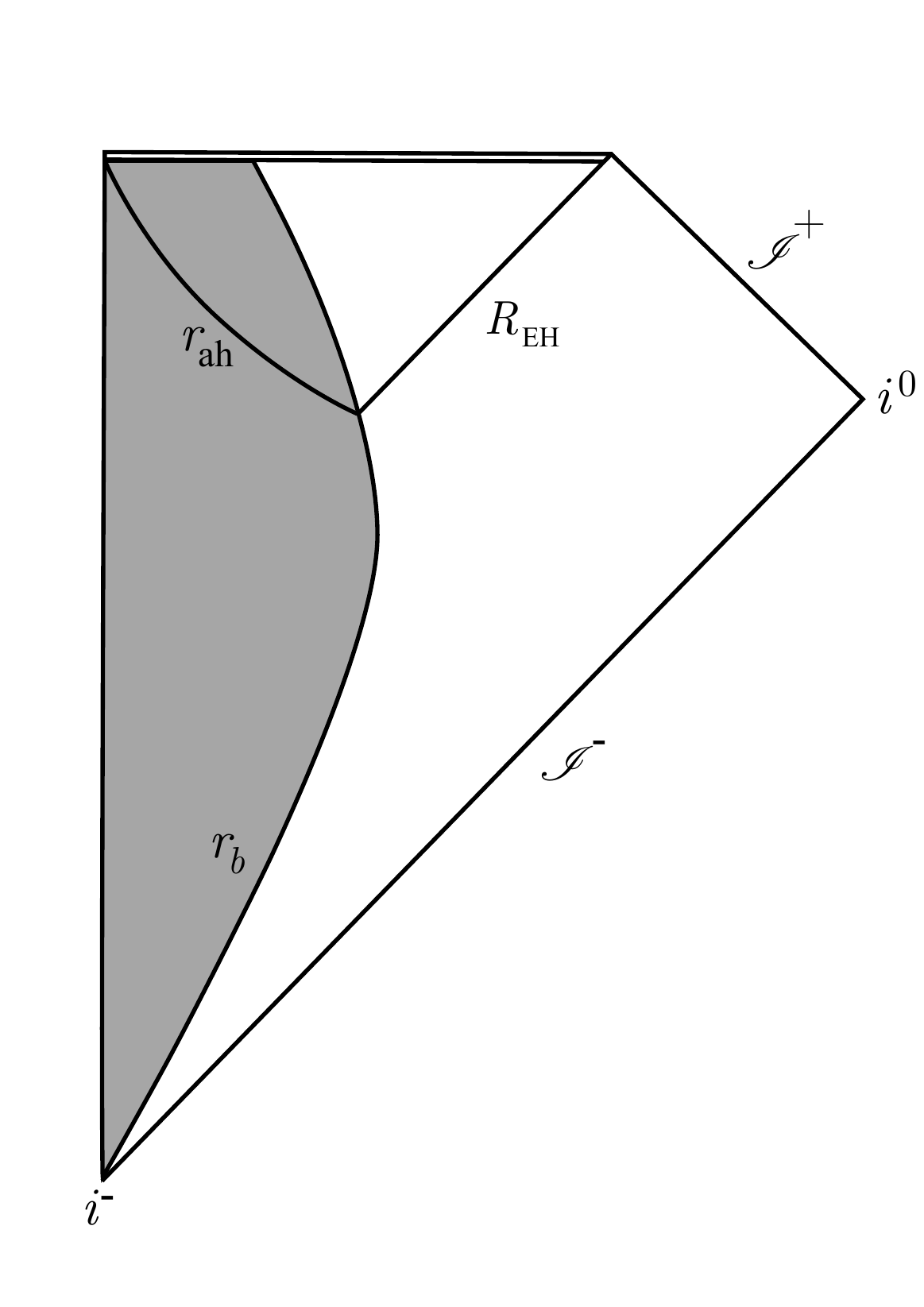}\quad \quad \quad \quad \quad
    \includegraphics[width=0.29\linewidth]{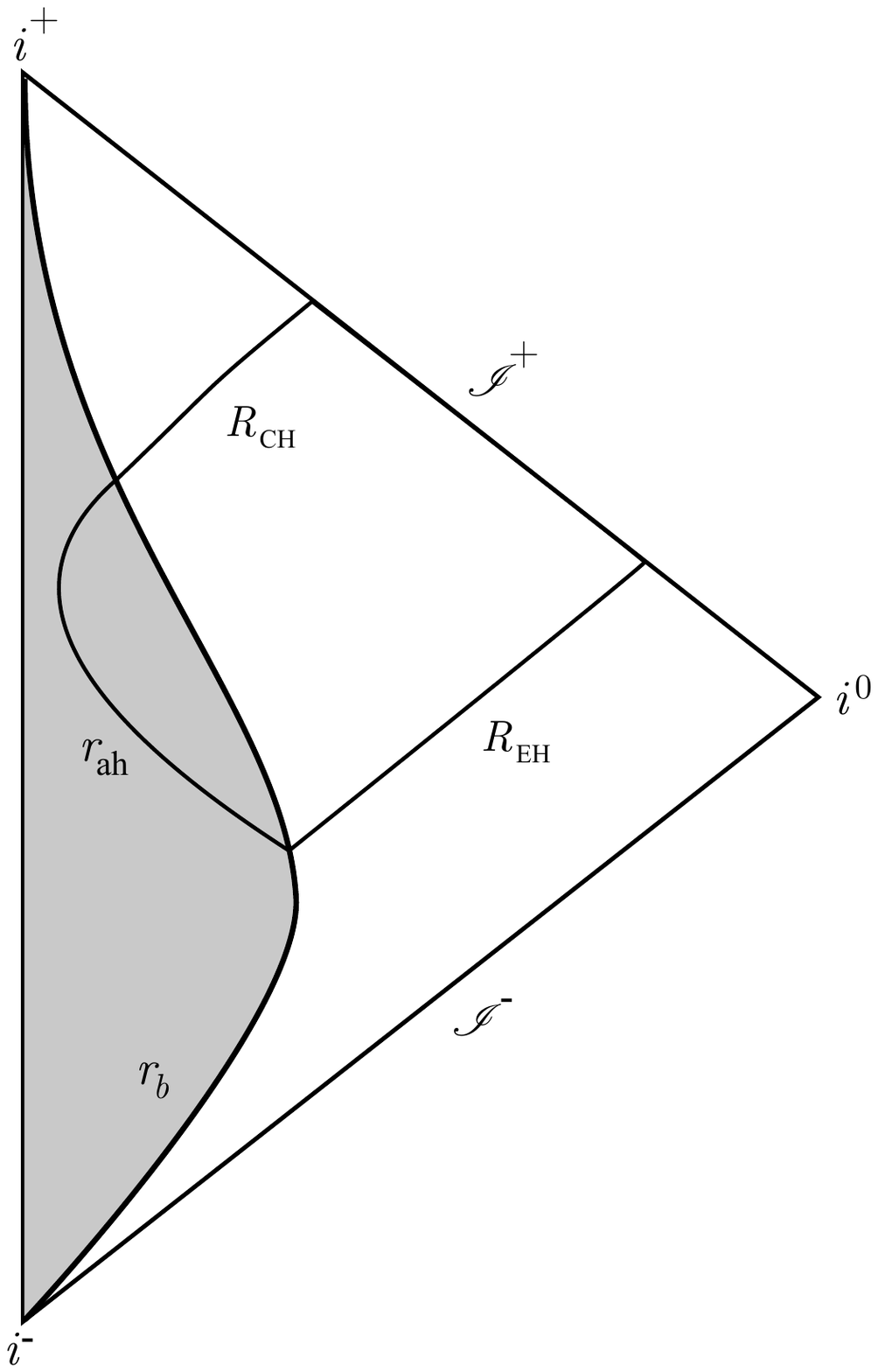} \quad \quad \quad
    \includegraphics[width=0.29\linewidth]{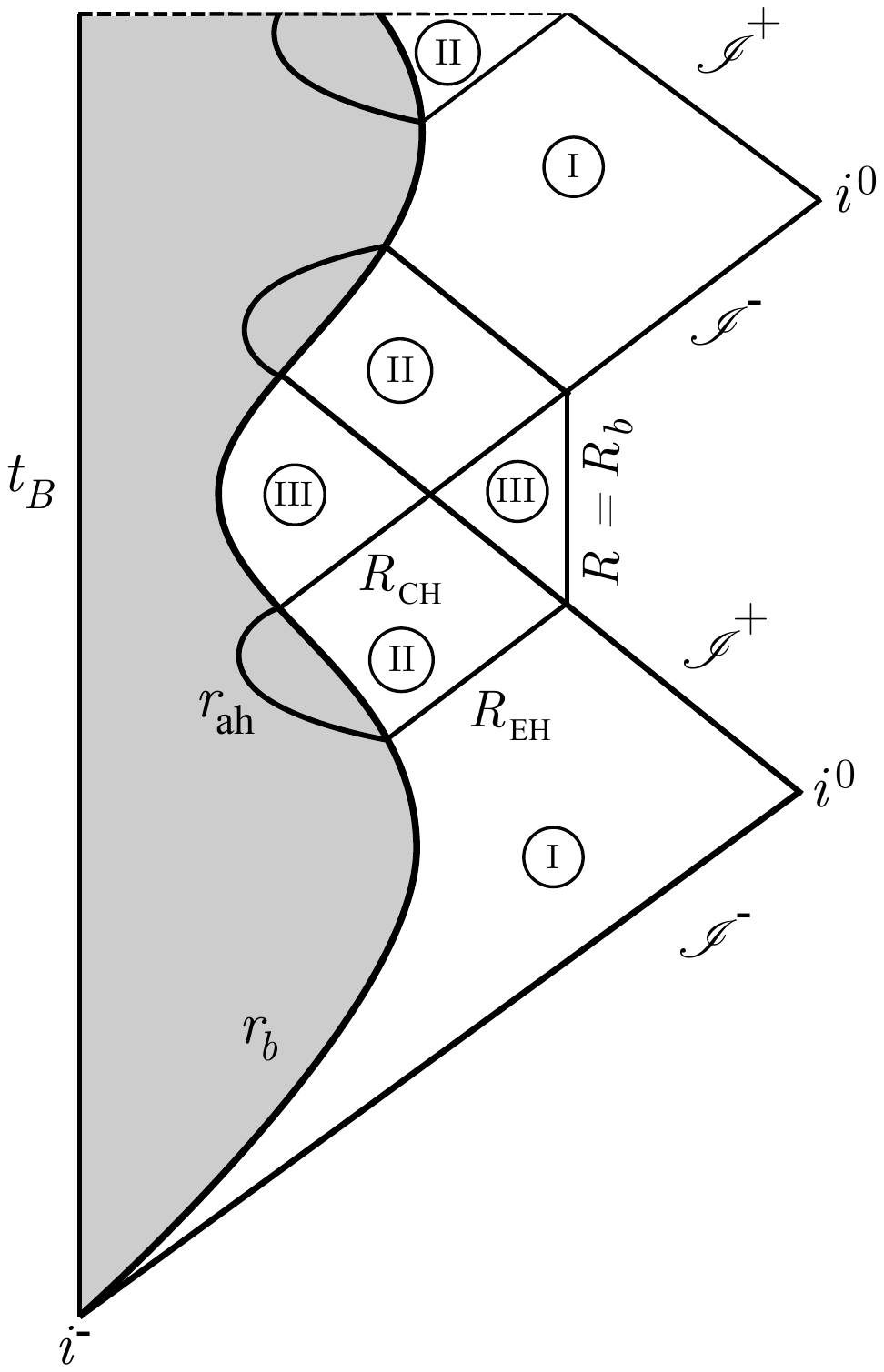}
\caption{Penrose diagram for the OSD collapse. The grey and white regions represents, respectively, the interior and exterior spacetime (left panel). Possible Penrose diagram for the non-singular marginally bound (central panel) and bound (right panel) collapse, for $M_{0}>M_{0_{\text{cr}}}$. Unlike the trajectory $r_{b}$ of the collapsing boundary, the trajectory $r_{\text{ah}}$ is not a world-line, but rather a mathematical curve indicating that a certain spherical shell layering the matter ball becomes a trapped (an untrapped) surface. The formation of the inner horizon is a direct product of the non-singular dynamics, and affects the global structure of spacetime. The construction of the diagram for the bounded case involves the use of Kruskal-Szekeres coordinates, and potentially gives rise to nested universes, traced by the periodic occurrence of the spacetime region dubbed with Roman number “I”.} \label{fig_penrosediagrams}
\end{figure*}

The matching of the extrinsic curvature, an object which describes how a three-dimensional hypersurface is influenced by the fact that it is immersed in a four-dimensional manifold, determines uniquely the mass function of the exterior (see Appendix \ref{Appendix1}). Taking into account \eqref{radialcoordinatematching}, it is enlightening to express $M(R)$ with respect to the potential appearing in the interior solution
\be\label{exteriomisner}
M(R; \xi, r_{b}, m_{0})=-\frac{{r_{b}}^2 R}{2} \, V\left(R; \xi, r_{b}, m_{0} \right) \, .
\ee
The latter entails the ending state geometry generated by the gravitational collapse of the matter ball. Notably, the functional form is independent of the value of $K$, and, in general, the solution for $a(t)$ is not needed to obtain the above expression. Moreover, it contains the information about the mass parameter $M_0$ of the exterior, which, as anticipated, is entirely constrained by the parameters of the interior:
\be
M_{0}\equiv \lim_{R \to \infty} M(R)=\lim_{\xi \to 0} M(\xi)=\frac{m_{0}r_{b}^3}{2} \, \, .
\ee
Thus, the non-trivial result that the limit to radial infinity and the limit - in parameters space - to GR do commute, provides an unambiguous definition for the mass of the regular BH, or of the horizonless BH mimicker, as a result of its dynamical formation. Indeed, it coincides with that of the star which, by collapsing, generates the non-singular spacetime\footnote{In particular, restoring the dimensional factor $8 \pi G_{\text{N}}$, the physical mass is $\mathcal{M}=M_{0}/G_{\text{N}}=m_{0}r_{b}^3/2G_{\text{N}}=(8\pi G_{\text{N}}\epsilon_0/3)r_{b}^3/2G_{\text{N}}=\left( \frac{4}{3}\pi r_{b}^3 \right) \epsilon_{0}$. Indeed, $m_0$ has dimension of a mass squared.}.  

The causal structure resulting from the collapse matches the evolution of the causal structure in the interior, and is described by the roots - and their degeneracy - of the function $\nabla_{\mu}R\nabla^{\mu}R$, given by the equation
\be
f(R; \xi, r_{b}, m_{0})=0 \, \, .
\ee
The latter one, for geometries generated by a non-singular dynamics with gravitational evanescence of collapsing matter, has two coincident solutions when $M_{0}$ equals the critical value $M_{0_{\text{cr}}}=M_{0_{\text{cr}}}(r_{{\text{ah}}_{\text{cr}}})$ (which occurs if and only if $r_b=r_{{\text{ah}}_{\text{cr}}}$), two different solutions for $M_{0}>M_{0_{\text{cr}}}$ ($r_b>r_{{\text{ah}}_{\text{cr}}}$), and no solutions for $M_{0}<M_{0_{\text{cr}}}$ ($r_b<r_{{\text{ah}}_{\text{cr}}}$). In the first case, an extremal BH with two coincident horizon forms. The second case represents the formation of a regular BH with an outer event horizon $R_{\text{EH}}$ and an inner Cauchy horizon $R_{\text{CH}}$; at the instants $T=T_{\text{EH}}(t_{\text{EH}})$ and $T=T_{\text{CH}}(t_{\text{CH}})$, when, respectively, the event and the Cauchy horizon form, the three-dimensional hypersurface $\Sigma$ is light-like, and at $t=t_{\text{CH}}$, the comoving shell $r=r_{b}$ returns to being untrapped. The third case represents a horizonless BH mimicker. The bijective bound between the final global causal structures and points in parameters space is depicted, for $K=0$, in Fig.\ \ref{fig_parameterscausal}. The phase diagram is qualitatively the same for the case $K=1$. Indeed, assigned a dynamics and a couple $(m_0, \xi)$, a different value of $K$ just reflects on the value of $r_{{\text{ah}}_{\text{cr}}}$. The Penrose diagrams in Fig.\ \ref{fig_penrosediagrams} visually summarize the whole collapse process at the level of the global spacetime solution\footnote{It is worth to mention that the construction of definitive Penrose diagrams involves highly non-trivial conceptual issues, like the interplay of the physics occurring in the region close to the bouncing matter with that at spatial infinity, and the replacement of teleological horizons with their dynamical and quasi-local counterparts. Such a discussion is out the scope of this work, and for clarifications on some of the issues at stake we refer the reader to \cite{Malafarina:2017csn, Han:2023wxg, Rovelli:2024sjl}.}. For a static external observer, the causal structures corresponding to the OSD model and to the non-singular bound and marginally bound models are qualitatively indistinguishable. However, for an observer comoving with the collapsing boundary, or for an external observer who decides to travel towards the inner region of the BH, they are not.

Finally, to fully characterize the exterior solution, it is crucial to introduce its Kretschmann scalar
\be
\begin{aligned}
\mathcal{K}(R) =& \frac{48M^{2}}{R^6}-\frac{64MM'}{R^5}+\frac{32(M')^{2}}{R^4} \\
  &+\frac{16M''}{R^4}-\frac{16M'M''}{R^3}\\
  &+\frac{4(M'')^2}{R^2} \, \, ,
\end{aligned}
\ee
where a comma represents a derivative with respect to the Schwarzschild-like radial coordinate, and the explicit form that the SEC assumes. Since the Einstein tensor can be reconstructed using the metric solution defined by \eqref{spacetimeexterior}, \eqref{lapseexterior} and \eqref{exteriomisner}, and since the field equations $G^{\mu}_{\, \, \nu}=T^{\mu}_{\, \, \nu}$ hold, with an anisotropic fluid $T^{\mu}_{\, \, \nu}=\text{diag}[-E, P_{\parallel}, P_{\perp}, P_{\perp}]$ sourcing the spacetime, the SEC can be expressed as
\be \label{exteriorsec}
E(R)+P_{\parallel}(R)+2P_{\perp}(R)=2P_{\perp}(R) \geq 0 \, \, ,
\ee
where $P_{\parallel}$ and $P_{\perp}$ are, respectively, the tangential and perpendicular pressure.

\section{Non-Singular Dynamics and Core} \label{Nonsingularcores}
A specific non-singular dynamics $\chi(\epsilon; \xi)$ determines uniquely the form of the solutions of $a(t)$ and $M(R)$ which violate the SEC. Thus, it determines the type of asymptotic core generated, manifested by the expansion of $M(R)$ in the small $R$ regime. According to condition \eqref{exteriorsec}, the violation of the SEC can occur by $P_{\perp}(R)$ approaching a negative finite value, as in the case of a de Sitter core, or zero from below, as in the case of a Minkowski core. Conversely, given a non-singular spacetime $M(R)$, it can be found out, inverting \eqref{exteriomisner} and \eqref{potential}, which one is the underlying dynamics for the gravitational collapse. Such non-singular scenarios fall always into one of the three following cases:
\begin{subequations}\label{eq:geometries}
\begin{align} \label{de Sitter}
&\lim_{\epsilon \to \infty} \Lambda(\epsilon) = c_{\xi}  & \iff \, &\mathrm{de \, \, Sitter \, \, ,} \\ \label{Minkowski}
&\lim_{\epsilon \to \infty} \Lambda(\epsilon) = 0  & \iff \, &\mathrm{Minkowski \, \, ,} \\ \label{SteepPressure}
&\lim_{\epsilon \to \infty} \Lambda(\epsilon) = \infty  & \iff \, &\mathrm{Steep \, \, Pressure \, \, ,}
\end{align}
\end{subequations}
where $c_{\xi}>0$ is a constant depending on $\xi$.

Although our framework and formalism are no more than effective, they hint on a fundamental possible interpretation of the different cores. While the matter ball is contracting, what remains limited is the quantity $G(\epsilon)T_{0}^{0 \, (m)}$: we have $G(\epsilon)\epsilon \to 0$ as $ \epsilon \to \infty$. At the same time it holds $[T_{0}^{0 \, (\text{eff})}-\Lambda(\epsilon)] \to 0$, where, in the Minkowski case, $G(\epsilon)\epsilon$ and $\Lambda(\epsilon)$ vanish separately instead of cross-cancelling. Thus, the number of particles is conserved - particles do not vanish in the process of collapse, as guaranteed by the mass-continuity equation, while the flow of the effective energy-momentum tensor characterises the transition to a vacuum-like state: it occurs a transition from a macroscopic Friedmann spacetime to a microscopic spacetime corresponding to one of the core types. All core types imply that the concept of mass of a particle depends on the energy scale, and that particles do not gravitate anymore when in the effective-vacuum state. On the other hand, each black-hole core may correspond to a different quantum-gravitational vacuum with its own defining symmetry, which in our effective framework is encoded in $\Lambda(\epsilon)$ and $P_{\perp}(R)$. The way the SEC is violated may be related to different stability properties of each vacuum state.

In the next subsection, we enlighten the general bound between the production of geometries with Minkowski core, as ending state of the collapse, and dynamics in which the nature of gravity turns intrinsically repulsive - due to $G(\epsilon)$ changing its signature. Then, we proceed with a case by case study, where we clarify the nomenclature for the third type of core, rooted in the peculiar behaviour of $p_{\text{eff}}(\epsilon)$ and $P_{\perp}(R)$.

\subsection*{General Feature of the Minkowski Core}

A Minkowski core via gravitational evanescence of collapsing matter is obtained for $ \Lambda(\epsilon \to \infty)=0$, in addition to \eqref{disappearance}. At the same time, we recall that it holds, like for every core, $\chi(\epsilon \to 0) \approx 1$, $G(\epsilon\to0)=1$, $\Lambda(\epsilon \to0)=0$, and $\Lambda(\epsilon) \neq \Lambda_{0}$ for $\epsilon \in \, ]0, \infty[$, with $\Lambda_{0}$ constant. It follows immediately that the formation of a Minkowski core requires $\Lambda(\epsilon)$ to be non-monotonic in $\epsilon \in \, ]0, \infty[$.
\\
\\
\textit{Property}:
\\
In general, each change $j$ of monotonicity for $\Lambda(\epsilon)$ occurring at $\bar{\epsilon}_{j}$, characterised by $d\Lambda/d\epsilon|_{\epsilon=\bar{\epsilon}_{j}}=0$, corresponds to a change of monotonicity for $G(\epsilon)$ at the same point, and viceversa. Other unpaired monotonicity variations are not allowed. Furthermore, if $\Lambda(\bar{\epsilon}_j)$ is a maximum (minimum), then $G(\bar{\epsilon}_j)$ is a minimum (maximum), and viceversa.
\\
\\
\textit{Derivation}:
\\
Taking the derivative of equation \eqref{G00} both sides, and replacing on its right-hand side the definition of $G(\epsilon)$ in terms of $\chi(\epsilon)$, we obtain $\epsilon(d G/d\epsilon)=-d\Lambda/d\epsilon$. Taking twice the derivative of equation \eqref{G00} both sides, we obtain $\epsilon(d^2 G/d\epsilon^2)+dG/d\epsilon=-d^2\Lambda/d\epsilon^2$. Then it is enough to evaluate the formers at $\epsilon=\bar{\epsilon}_j$.
\\

At this point, we have all the ingredients to prove the general feature discriminating the dynamics sourcing a spacetime with Minkowski core.
\\
\\
\textit{Hypotheses}:
\begin{enumerate}[label=(\roman*)]
\item $\chi \in C^\infty(\epsilon)$ and $\chi \geq 0$ for $\epsilon \in [0, \infty[$
\item $j=1$, \textit{i.e.}\ there is only one change of monotonicity
\item Matter is homogeneous, and $p(\epsilon) \geq 0$, \textit{i.e.}\ exotic matter is excluded
\item The collapse generates a geometry with Minkowski core.
\end{enumerate}
\begin{space}
\end{space}
\begin{space}
\end{space}
\textit{Thesis}\footnote{Assumptions (i) and (ii) are not that restrictive and just exclude bizarre dynamics that would be difficult to justify physically. Indeed, $\chi(\epsilon)<0$ or $j>1$ would imply, at intermediate stages of the collapse, multiple oscillations for the character of gravity and the potential.}:
\\
At finite energy-density values, before vanishing asymptotically, $G$ assumes negative values.
\\
\\
\textit{Proof}:
\\
As a consequence of (ii), only two cases are possible.
\begin{enumerate}[label=(\alph{enumi})]
\item $\Lambda(\epsilon)>0$ for $\epsilon \in \, ]0, \infty[$. It follows that $\Lambda(\bar{\epsilon})>0$ is a maximum. Due to the aforementioned property, $G(\bar{\epsilon})$ is a minimum. Moreover, it must hold $G(\bar{\epsilon})<0$. Indeed, if it was $G(\bar{\epsilon})>0$, since it is $G(\epsilon \to 0)=1$ and $j=1$, it would be impossible to have $\lim_{\epsilon \to \infty}G(\epsilon)=0$. Thus, since $G(\epsilon \to 0)=1$ and $G(\bar{\epsilon})<0$, there exists a point $\epsilon_{*}$ such that $\epsilon_{*}<\bar{\epsilon}$ and $G(\epsilon_{*})=0$, and it holds $G(\epsilon)<0$ for all finite values $\epsilon > \epsilon_{*}$
\item $\Lambda(\epsilon)<0$ for $\epsilon \in \, ]0, \infty[$. It follows that $\Lambda(\bar{\epsilon})<0$ is a minimum and $G(\bar{\epsilon})$ a maximum. It must hold $G(\bar{\epsilon})>0$ and, at the same time, $G(\epsilon)>0$ for all finite $\epsilon>0$, because, otherwise, it would be impossible to have $\lim_{\epsilon \to \infty}G(\epsilon)=0$. However, if for $\epsilon \in \, ]0, \infty[$ we have $\Lambda(\epsilon)<0$, $G(\epsilon)>0$, and, as required by (i), $\chi(\epsilon)>0$, then the effective energy-momentum tensor with a fluid satisfying (iii) will not violate the SEC, as a quick look at \eqref{interiorSEC1}-\eqref{interiorSEC3} reveals: this is \textit{ad absurdum} with (iv), indeed, due to the singularity theorem, it would not be possible to avoid the singularity formation and generate a regular core of any type, including - of course - the Minkowski one.
\end{enumerate}
\begin{space}
\end{space}
\begin{space}
\end{space}
One has to conclude that it holds (a), and the thesis is demonstrated. The general feature encapsulated in this theorem is the main result of our work. In our opinion, the most interesting aspect relies in the novel and non-trivial way in which the energy condition, the formation of a non-singular spacetime, and the running of the couplings are just intertwined.
\begin{figure*}[htbp!]
\centering
    \includegraphics[width=0.5\linewidth]{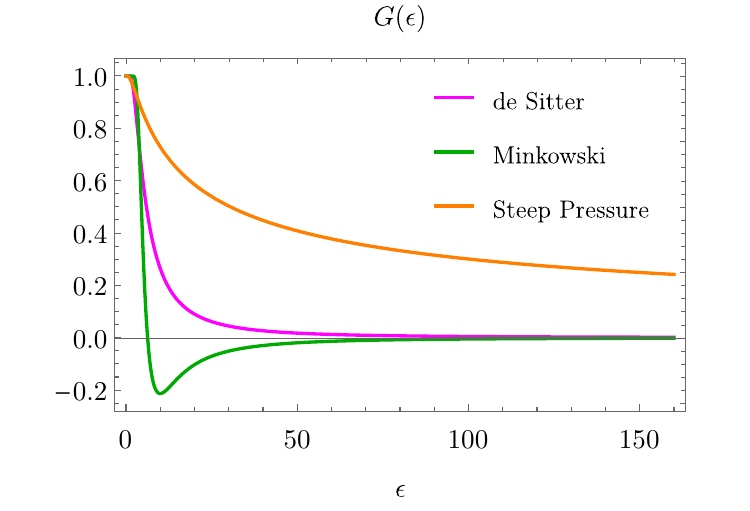}\hfil \hfil 
    \includegraphics[width=0.485\linewidth]{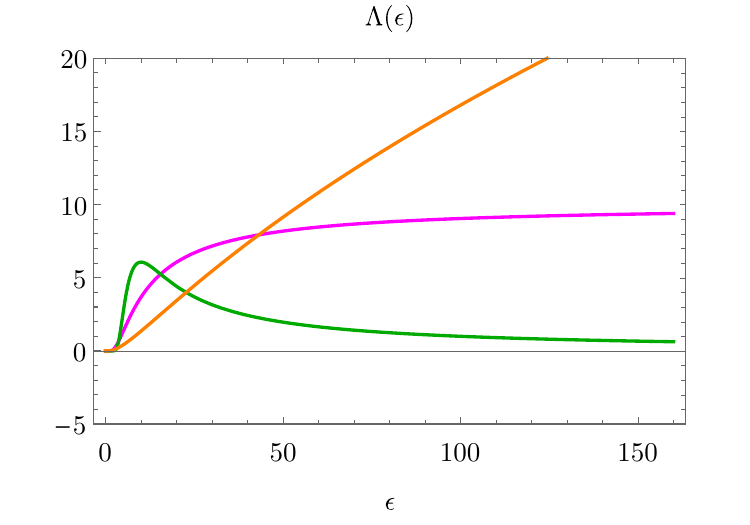}\par\medskip
\caption{Dependence of Newton coupling (left panel) and cosmological coupling (right panel) on matter energy-density, for the three dynamics studied, for $\xi=0.1$. In the dynamics generating the non-singular spacetime with Minkowski core, $G(\epsilon)$ assumes negative values before vanishing and has a negative minimum when $\Lambda(\epsilon)$ has a positive maximum. Here, for illustrative purpose, $\epsilon$ follows a linear scale, instead of solving \eqref{dust_energy_density} with an actual solution $a(t)$.} \label{fig_G_L}
\end{figure*}

\begin{figure*}[htbp!]
\centering
    \includegraphics[width=0.495\linewidth]{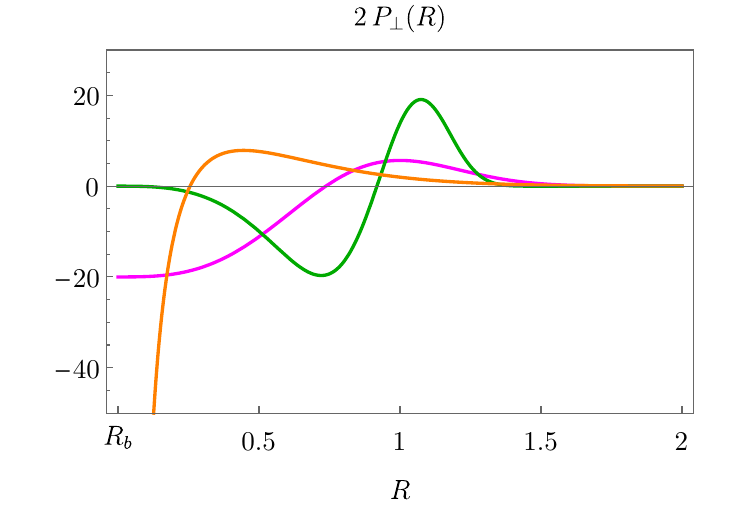}\hfil \hfil 
    \includegraphics[width=0.495\linewidth]{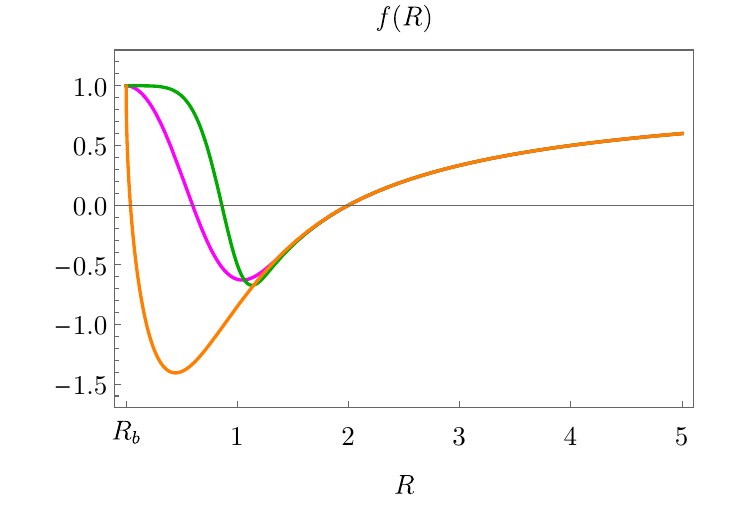}\par\medskip
\caption{Perpendicular pressure of the anisotropic fluid sourcing the exterior spacetime (left panel) and behaviour of the exterior metric component in the case $M_0>M_{0_{\text{cr}}}$ (right panel), for $\xi=0.1$, $M_0=1$. In the non-singular spacetime with steep pressure core, $P_{\perp}(R)$ has one inflection point only, thus tends to diverge negatively in the small $R$ regime.} \label{fig_P_f}
\end{figure*}


\subsection*{Case Study Dynamics}
In order to further categorise and compare the three different cores, we propose an explicit dynamics $\chi(\epsilon; \xi, p)$, where $p$ is a new dimensionless parameter. The latter controls the rate at which gravity weakens, allowing us to discuss at least one non-singular solution for each type of core. Clearly, $p$ can not be absorbed into $\xi$, nor into one of the other parameters. Thus, once assumed that $\xi$ has a non-zero value, there is still plenty of freedom in determining the functional form of $\chi(\epsilon)$. Different forms can lead to sensibly different physics. However, we expect that if the non-singular spacetime is realized in nature, it has always the same type of core, and the collapsing dynamics responds always to the same physical law. In this sense, here the parameter $p$ only represents our ignorance of which non-singular dynamics is preferred by nature. Similarly to the issue of the primary origin of $\xi$, we expect that a more fundamental theory could shed light on this aspect.

We choose to work with a generalization of the dynamics sourcing the Dymnikova black hole \cite{Dymnikova:1992ux, Dymnikova:2001fb}, namely
\be \label{dynChip}
\chi(\epsilon)=1- e^{-\frac{1}{(p-1)\xi^{p-1} \epsilon^{p-1}}} \, \, , \, \, p>4/3 \, \, ,
\ee
where the lower bound for $p$ is fixed by requiring, a posteriori, the regularity of $V(a)$ at all $a$; in particular, the case $p=4/3$ gives a smooth potential but does not satisfy condition \eqref{regularpotential}.  Indeed, in our framework, the Dymnikova black hole is generated by the non-singular dynamics corresponding to $p=2$. The straightforward generalization to other values of the parameter corresponds to implement a different cutoff in the effective energy-density $\epsilon_{\text{eff}}=\chi(\epsilon)\epsilon$. This clarifies to which physical feature is related $p$ in general. The energy-density cutoff, as tool to obtain non-singular collapse models, is widely adopted in the literature, and it is usually implemented at the level of the field equations only \cite{Barcelo:2007yk, Bambi:2013caa, Liu:2014kra, Malafarina:2022oka}, where is encoded within the effective energy-momentum tensor. Instead, in our framework, it is additionally constrained within the action and condition of gravitational evanescence of collapsing matter\footnote{It is worth to observe that \eqref{dynChip} can be obtained as solution of the differential equation
\be \label{diffeqdynChip}
\chi-\xi^{p-1}\frac{\partial \chi}{\partial \epsilon}\epsilon^{p}=1 \, \, ,
\ee
and fixing the integration constant to $C=-1$ by requiring, a posteriori, the regularity of $V(a)$ at all $a$. Furthermore, a regular black hole of the Dymnikova class is found in \cite{Platania:2019kyx}, applying a “dynamical renormalization” of the Schwarzschild vacuum. Such a procedure leads to a differential equation, namely eq.\ (22) in \cite{Platania:2019kyx}, whose solution is a varying Newton constant for the exterior geometry. Thus, there is a complementarity between the latter and \eqref{diffeqdynChip} with $p=2$, which primary refers to the collapsing matter interior.}. Finally, the dynamics \eqref{dynChip} satisfies all the hypotheses assumed in the previous theorem.

\subsubsection*{de Sitter}
As anticipated, for $p=2$, it occurs a dynamics with asymptotically finite and positive cosmological constant. The scale factor at large times, for the case $K=0$, and the resulting non-singular spacetime are, respectively,
\begin{subequations}\label{soldeSitter}
\begin{align} \label{deSitter_scalefactor}
a(t) &\sim e^{-t/\sqrt{3\xi}} \, \, , \, \, t \to \infty \, \, , \\ \label{deSitter_mass}
M(R)&=M_{0}\left(1-e^{-\frac{R^3}{6 \xi M_{0}}}\right) \, \, .
\end{align}
\end{subequations}
For the case $K=1$, $a(t)$ can be obtained only numerically. The behaviour of the varying couplings, and of the perpendicular pressure and metric component, are depicted, respectively, in Fig.\ \ref{fig_G_L} and Fig.\ \ref{fig_P_f}. In Tab.\ \ref{tab:corestresults} we report the general features of the de Sitter core. They are independent from the dynamics we have chosen. In particular, expanding around\footnote{This choice of $R_0$ is due to better readability of expressions and to consistency with previous definitions of “core” in the literature \cite{Simpson:2019mud}. However, where needed, we shall remark that in our model the result of the expansion can not be extrapolated up to $R<R_b$, where $R_b>0$, if one is looking for the correct physical interpretation.} $R_0=0$, the exterior mass function, the perpendicular pressure and the Kretschmann scalar evaluate, respectively, to
\begin{subequations}\label{deSitterfeatures}
\begin{align}
\label{massexpansion}
M(R)&=\frac{R^3}{6\xi}+O(R^6) \, \, , \\
\label{pressureexpansion}
2P_{\perp}(R)&=-\frac{2}{\xi}+\frac{5 R^3}{6 \xi^2 M_0}+O(R^6) \, \, , \\
\label{kretschmannexpansion}
\mathcal{K}(R)&=\frac{8}{3\xi^2}+O(R^3) \, \, ,
\end{align}
\end{subequations}
as expected for a de Sitter core. 


\subsubsection*{Minkowski}
Taking $p=3$, we can investigate the consequences of increasing the rate at which gravity weakens, due to a faster cutoff in the energy-density. The qualitative behaviour of $G(\epsilon)$ and $\Lambda(\epsilon)$ changes significantly too, as depicted Fig.\ \ref{fig_G_L} and Fig.\ \ref{fig_P_f}. Notably, in this case, the SEC in the interior is violated already via \eqref{interiorSEC2}, due to the Newton constant becoming negative, in agreement with what we have just proved as general feature of the Minkowski core, while the dynamics generating the de Sitter core violates only \eqref{interiorSEC3}. Although the condition $\Lambda(\epsilon)/\epsilon=\chi(\epsilon) \neq 0$ is general in form, the finite value $\epsilon_*$ at which holds $G=0$ depends both on the underlying fundamental dynamics $\chi(\epsilon; \xi)$ and, through \eqref{dust_energy_density} and the scale factor solution, on the rest of the parameters space. In particular, in order to avoid inconsistencies with the early stage of the collapse regimented by GR, it should hold $\epsilon_{*}\gg \epsilon_{0}$.

\begin{table*}[htbp!]
\caption{\label{tab:corestresults} General properties defining the different cores that can be obtained by gravitational evanescence of collapsing matter. These properties hold beyond and regardless of our specific case study.}
\begin{ruledtabular}
\begin{tabular}{l||c|c|c}
\textbf{Core} & \textit{\textbf{de Sitter}} & \textit{\textbf{Minkowski}} & \textit{\textbf{Steep Pressure}}  \\ \hline \hline
$G(\epsilon)\geq 0 \, \, \, \, \forall \, \epsilon$ & \checkmark & \crossmark & \checkmark \\ \hline
$d \Lambda(\epsilon)/d \epsilon > 0 \, \, \, \, \forall \, \epsilon$ & \checkmark  & \crossmark & \checkmark \\ \hline
$\lim_{\epsilon \to \infty}\Lambda(\epsilon)$ & $c_{\xi}$ & $0$ & $+ \infty$ \\ \hline
$\frac{d^2P_{\perp}(R)}{dR^2}\big|_{R=R_l}=0 \, \, , \, \, l \in \mathbb{N} \, \, : \, \,  l>1$ & \checkmark & \checkmark & \crossmark \\
\end{tabular}
\end{ruledtabular}
\end{table*}

For the aforementioned value of $p$, the interior and exterior solutions are, respectively,
\begin{subequations}\label{solMinkowski}
\begin{align} \label{deSitter_scalefactor}
a(t) & \sim \left( \frac{8 \xi^2 \epsilon_0}{3t^2} \right)^{1/3} \, \, , \, \, t \to \infty \, \, , \\ \label{deSitter_mass}
M(R)& =M_{0}\left[1-e^{-\frac{R^6}{2 (6 \xi M_{0})^2}}\right] \, \, .
\end{align}
\end{subequations}
Thus, in this case the scale factor scales as a polynomial. The innermost region of the exterior spacetime is characterised by
\begin{subequations}\label{Minkowskifeatures}
\begin{align}
\label{massexpansion}
M(R)&= \frac{R^6}{72 \xi^2 M_0}+O(R^{12}) \, \, , \\
\label{pressureexpansion}
2P_{\perp}(R)&= -\frac{5 R^3}{6 \xi^2 M_0}+O(R^9) \, \, , \\
\label{kretschmannexpansion}
\mathcal{K}(R)&= \frac{7R^6}{18\xi^4M_0^2}+O(R^{12}) \, \, .
\end{align}
\end{subequations}
All the dynamics obtained for $p>2$ generate a Minkowski core, whose general defining features are reported in Tab.\ \ref{tab:corestresults}, and we have singled out the solution with $p=3$ just for the sake of simplicity. While a de Sitter core is generated exclusively by the dynamics corresponding to $p=2$.


\subsubsection*{Steep Pressure}
For $4/3<p<2$, the cosmological constant diverges in the asymptotic regime, and the geometry resulting from the collapse has a steep pressure core. For instance, taking $p=3/2$, we have the solution
\be
\label{steep_mass}
M(R) = M_{0} \left[ 1-e^{-\frac{2 R^{3/2}}{(6 \xi M_0)^{1/2}}} \right] \, \, .
\ee
In this case it is not possible to give an analytical expression for $a(t)$, because in the expansion of the potential the terms following the first are relevant. Asymptotically close to the surface of the matter ball, we have
\begin{subequations}\label{SteepPressurefeatures}
\begin{align}
\label{massexpansion}
M(R)&= q_1R^{3/2} + O(R^3)\, \, , \\
\label{pressureexpansion}
2P_{\perp}(R)&=-\frac{w_1}{R^{3/2}}+\frac{4}{\xi} + w_2 R^{3/2} +O(R^3)\, \, , \\
\label{kretschmannexpansion}
\mathcal{K}(R)&=\frac{z_1}{R^3} -\frac{z_2}{R^{3/2}}+\frac{143}{9\xi^2}+O(R^{3/2}) \, \, ,
\end{align}
\end{subequations}
where $q_1$, $w_1$, $w_2$, $z_1$, and $z_2$ are positive coefficients depending on $\xi$ and $M_0$. The dynamics generating this type of core violate the SEC only via \eqref{interiorSEC3}, similarly to de Sitter case. On the other hand, for the steep pressure core, while $M(R)$ converges to zero for $R=0$, the curvature diverges. However, \textit{it is physically meaningless to evaluate at $R=0$ quantities related the exterior spacetime}, since for the latter the domain of the radial coordinate is dictated by \eqref{domainradial}. In order to evaluate correctly the curvature in the center, as in any inner layer with respect to the surface shell constituting the boundary of the matter ball, it has to be considered the interior spacetime. Thus, it has to be used the expression \eqref{interiorKretschmann}. The latter, since $a(t)>0$ always, is finite at all times. However, we are left with a non-trivial question: given that the SEC is violated both in the interior and in the exterior, what does cause this mathematical divergence? The answer lies in the behaviour of $\Lambda(\epsilon)$, which reflects on that of $P_{\perp}(R)$. Peculiarly, the collapsing matter avoids the singularity formation thanks to an asymptotically increasing $\Lambda(\epsilon)$, which provides an indefinite amount of negative effective pressure $p_{\text{eff}}(\epsilon)$ in the interior. In the exterior, this consistently reflects onto $P_{\perp}(R)$ being not bounded from below for decreasing values of $R$. Such a defining property can be formalized as $P_{\perp}(R)$ having one inflection point only, as indicated in Tab.\ \ref{tab:corestresults}.


\section{Conclusion and Outlook}\label{Conclusion}
\textit{Summary}.\ The study of the dynamical process of collapse, within frameworks that, in one way or another, deviate from GR, provides additional physical insights into the issue of singularity resolution \cite{Malafarina:2024qdz}, as compared to considering static situations only. Furthermore, since the OSD model has played an important historical and scientific role in determining the physical relevance of BHs in GR \cite{Marcia, PenroseInterview}, obtaining non-singular space-time solutions as the final product of gravitational collapse reinforces their status as legitimate astrophysical candidate as an alternative to singular BHs. The alternative model presented in this work has a solid physical interpretation that can be summarized into the following steps:
\begin{enumerate}
\item When all thermonuclear sources of energy are exhausted a sufficiently heavy star will collapse. The first stages of the collapse are described accordingly to GR: the interior is modelized by a presureless FLRW solution, and the exterior by a Schwarzschild vacuum with $R_{b}>2M_{0}$. If $r_b>r_{{\text{ah}}_{\text{cr}}}$, the formation of the event horizon, occuring for $R_{b}= r_{b}a(t_{\text{EH}})=R_{\text{EH}}(\xi, M_0)\approx 2M_{0}$, can be considered almost as a consequence of the dynamics of GR only. However, \textit{the event horizon of a regular BH of mass $M_{0}$ will never be located at exactly the same position as that of a singular BH of same mass}.
\item When a certain energy-density threshold, whose value is related to $\xi$, is reached, the deviation from GR becomes relevant, and \textit{the potential perceived by the matter ball becomes repulsive}. The exterior spacetime in the region closer to $R_{b}$ is not well-described by a Schwarzschild vacuum with $R_{b}<2M_{0}$ anymore, but rather by a generalized Schwarzschild solution, whose specific form of $M(R)$ depends on $\chi$. The repulsive effect, caused by the asymptotical disappearance of gravitational interaction of collapsing matter, allows for a violation of the SEC. Thus, \textit{the singularity theorem can be circumvented, both in the interior and exterior spacetime}.
\item If $r_b>r_{{\text{ah}}_{\text{cr}}}$, at a later time, the inner horizon forms as a further consequence of such a repulsive effect, as the Penrose diagram makes manifest. In any case, regardless of the presence of horizons and of the dynamics, \textit{the energy of the collapsing dust matter of the interior is partly stored in the gravitational field of the non-singular global spacetime, being effectively transferred via the non-minimal coupling}. This mechanism seems a key aspect in the formation of non-singular spacetime, and it may be a key ingredient in quantum gravity as well. Indeed, the black-hole core may be interpreted as a vacuum state of non-gravitating particles, whose symmetry and stability are encoded in $\Lambda$ or $P_{\perp}$.
\item Lastly, the singularity formation is avoided in one of the two following ways, according to the initial energy of the particles composing the star.
\begin{enumerate}
\item For $K=0$, despite the repulsive potential, the ball keeps contracting forever, without never degenerating into a point mass. The \textit{geometric density} $1/a^3$, and the matter energy-density, keep increasing but remain finite at all times.
\item For $K=1$, the ball faces a bounce. At the instant of the bounce the potential starts to fuel the expansion of the ball, until the minimum of the potential is approached from the left. Then the perceived force becomes attractive again, to the point that the expansion is halted and the ball undergoes a crunch. At the instant of the crunch the whole process restarts in a nested universe.
\end{enumerate}
\end{enumerate}
We have shown that the dynamics leaves a permanent imprint on the innermost part of the exterior geometry, as encoded in \eqref{potential} and \eqref{exteriomisner}. Our study establishes that \textit{the speed of the effective energy-density cutoff determines the type of core generated}. Through a dynamics embedding a cutoff corresponding to a value $p>2$, we obtained a new Minkowski geometry. To the best of our knowledge, this is the first time that a spacetime with a Minkowski core is obtained via the dynamical process of gravitational collapse of a matter source, although the latter is highly idealised.
\\
\\
\textit{Rethinking Current Literature}.\ Curiously, the regular BHs with Minkowski core proposed in \cite{Culetu:2014lca, Ghosh:2014pba, Simpson:2019mud}, feature an exponential functional form too, but the mass function converges to zero skipping polynomial terms of any order, in particular as an exponential. Moreover, it is straightforward to observe that the spacetime with logarithmic mass function originally found in \cite{Bonanno:2023rzk} has a steep pressure core, similarly to our case for $4/3<p<2$. It is less immediate, and perhaps more surprising, to observe that the solution found in \cite{Borissova:2022mgd} - see equation (6.1) in the same reference - also has a steep pressure core, coinciding with that generated by our dynamics with $p=3/2$. 

Furthermore, in light of the theorem we have proved, it is suggestive to note that among the various regular BHs obtained in the literature within the Asymptotic Safety scenario, where $G(k) \geq 0$ is known to always hold, there is none which exhibits a Minkowski core (see, \textit{e.g.,}\ \cite{Bonanno:2000ep, Platania:2019kyx, Borissova:2022mgd, Bonanno:2023rzk, Harada:2025cwd, Bonanno:2025dry} for a partial but significant list). From the point of view of Asymptotic Safety, the condition $G(k) \to 0$ as $k \to \infty$ stems from the presence of the Reuter fixed point. Moreover, the singularity resolution seems to be robust under various cutoff identifications $G(k) \mapsto G(\epsilon)$ \cite{Bonanno:2023rzk, Harada:2025cwd, Bonanno:2025dry}. However, a relevant issue left for future studies is to understand which cutoff identifications preserve the monotonicity of $G$ and $\Lambda$ and which ones alter it, discussing how the presence of a zero in $G(\epsilon)$ at finite $\epsilon$ could affect the beta function of the dimensionless coupling. This may allow to figure out whether a Minkowski core is fatally excluded by the Asymptotic Safety approach or not. For instance, the expression $G(k)=G_{\text{N}}/(1+ G_{\text{N}} k^2/g_*)$ determines either a steep pressure core, using the map in \cite{Bonanno:2023rzk}, or a de Sitter core, using that in \cite{Harada:2025cwd}. Anyway, our results stem from the more general condition \eqref{disappearance}, therefore, our categorization of the regular black-hole cores in Tab.\ \ref{tab:corestresults} remains always valid, regardless of this issue.

Further insights could be provided by the study of the trace mode in the context of functional renormalization group techniques implemented on Lorentzian foliated spacetime \cite{Saueressig:2023tfy, Korver:2024sam, Saueressig:2025ypi}: since the trace mode represents the variation of the three-dimensional \textit{geometric volume} \cite{Reuter:2019byg}, its quantum fluctuations might be relatable to the non-singular geometrical density in classically-modified models of collapse.
\\
\\
\textit{Applications and Strengths}.\ Notably, most of our results are analytical, which make them suitable to investigate phenomenological implications. Indeed, as a natural follow-up, it would be interesting to study and compare the thermodynamical and stability properties of solutions obtained in this work, as well as their charged or rotating counterparts, and cosmological models related to the dynamics \eqref{dynChip}. In the case of BHs sourced by a negative Newtonian constant, the study of these properties may provide further informations about consistency bounds in parameters space, and so about their astrophysical viability.

Remarkably, our modified theory, at least in the dust matter case, does not alter the non-linearity of GR. As a consequence, in contrast to what happens in other frameworks, our model effortlessly reduces to the OSD model by a limit in parameters space, and generates regular BH and horizonless BH mimicker solutions with a clear limit to the Schwarzschild solution. Moreover, within a fixed dynamics, the position of the theory in its parameters space and the causal structure of the non-singular geometries, are related by a map that remains always bijective. In particular, the extremal regular BHs lie on a two-dimensional surface of the phase diagram. Thus, we have shown that their formation as ending state of gravitational collapse is less likely to occur with respect to the formation of non-extremal configurations.
\\
\\
\textit{Limitations}.\ However, our approach does not come free of important limitations. In particular, we signal the conceptual and technical challenges to extend the framework to more sophisticated matter sources, as well as to non-homogeneous fluids, due to the theory being not manifestly covariant under general transformations of coordinates. This comes as a consequence of the peculiar appearance of the energy-density in the effective Lagrangian. Furthermore, it still remains not completely clear what sources exactly the exterior solution in our model. Promisingly, such an issue could be addressed within the new frameworks proposed in \cite{Boyanov:2025pes, Carballo-Rubio:2025ntd} and \cite{Zhang:2025ccx, Alonso-Bardaji:2025hda}, which seem to be able to encompass a large class of non-singular solutions as product, respectively, of a Horndeski theory \cite{Horndeski:1974wa, Kobayashi:2019hrl} and of covariant effective-Hamiltonian dynamics. Despite the current limitations, the Markov-Mukhanov approach is preferable to other proposals insofar as it allows to easily implement the concept of asymptotic freedom for a gravity-matter system. Here, the scale dependence hints intuitively at a specific unification of space, time \textit{and} matter \cite{Markov:1983}. Thus, we have prioritized principles instead of formalism.
\\
\\
\textit{Possible Universal Aspects}.\ Within our theory, all the possible collapsing dynamics generating a Minkowski core as ending state are related to a Newton constant assuming negative values. This conclusion comes from a theorem which binds up together all the key ingredients appearing in our work: the behaviour of the varying couplings, the way in which the SEC is violated, and the type of regular core generated. In particular, a negative $G$ amounts to a radical change in the nature of gravity, which would turn intrinsically unfocusing with respect to all the four dimensions. Massive bodies with positive energy-density would be repelled one from each other, sourcing a negative, locally hyperbolic, spacetime curvature. A change of sign may suggest an earlier transition of some kind, preceding the asymptotic one related to the final quantum-gravitational vacuum state. While the idea of a quantum-transition has already been employed to interpret the de Sitter core \cite{Sakharov:1966hcb, Hartle:1983ai, Poisson:1988wc, Frolov:1988vj, Dymnikova:1992ux, Dymnikova:2019cmb}, understanding how a couple of consequent transitions would be related to a Minkowski spacetime remains an open question. At the effective level of a modified gravity theory, ultimately, another question arises: is the theorem general even beyond the specific theory adopted here? In this regard, it might be relevant to check whether it holds for other modified theories with non-minimal gravity-matter coupling \cite{Dicke:1961gz, Hehl:1976kj, Trautman:2006fp, Westman:2014yca}. Intriguingly, a study on the equivalence between scale-dependent gravity and scalar-tensor theories has just appeared in the literature \cite{Neckam:2025kip}. However, we expect that covering this step will require to rely heavily on numerical integrations.

To conclude: why is it useful to address the issue of the degree of generality of this feature? First, in the speculative scenario in which the theorem is proven to hold universally, if deviations from GR relatable to a spacetime with asymptotically Minkowski core will be detected, any viable approach to quantum gravity should be able to encapsulate the constrained correlation on the behaviour of the Newton constant; conversely, if a certain fundamental theory, that gains the status of being the one that describes nature better, does not provide the feature of a negative constant \cite{Ayuso:2019bnw, Volovik:2022sdr}, one should focus on the study and the detection of deviations relatable to other types of core only. Second, our finding evokes the possibility of a broader picture where a geometrizable concept like that of energy condition - and its violation, crucial to figure out the ultimate fate of gravitational collapse, could be systematically related to that of running coupling, hallmark of the quantum theory of fields. It seems reasonable to advocate that any attempt to establish these kind of bonds may constitute a channel to learn new informations on the nature - if any - of quantum gravity. We hope to delve deeper into these issues in future researches.

\section*{Acknowledgements}
We are thankful to an anonymous referee for constructive comments. We are grateful to Alfio Bonanno, Diego Buccio and Daniele Malafarina for important discussions, to Jesse van der Duin for “a little help” in generating the three-dimensional plot in Fig.\ \ref{fig_parameterscausal}, and to Luca Buoninfante for a careful reading of the latest draft, as well as for perpetual meta-physical support. We are thankful to Patrick Bourg, Raúl Carballo Rubio, Badri Krishnan, Renate Loll, Alessia Platania, Frank Saueressig, Marc Schiffer and Vania Vellucci for extremely helpful feedback on the earlier version of the manuscript. This research has been partially funded by Erasmus+ grants no.\ 2022-1-IT02-KA131-HED-000066529 and no.\ 2023-1-IT02-KA131-HED-000135582.

\appendix

\section{Junction Conditions}\label{Appendix1}
The application of the junction conditions ensures that geodesics are continuous across the region where the discontinuity of the matter energy-momentum tensor localizes, and, thus, in all the spacetime manifold $\mathcal{M}$. In order to introduce the conditions, we follow closely the derivation as presented in \cite{Poisson:2009pwt, Joshi:2008zz}.

The global four-dimensional manifold can be divided in two portions, $\mathcal{M}=\mathcal{M}_{-}\cup \, \mathcal{M}_{+}$, separated by a \textit{three-dimensional boundary hypersurface $\Sigma$ following a time-like trajectory}. Since the field equations hold in both regions of spacetime, the line elements in $\mathcal{M}_{\pm}$ can be written as 
\be
ds^2_{\pm}=g_{\mu \nu \pm} dx^{\mu}_{\pm}dx^{\nu}_{\pm} \, \, .
\ee
The line element induced on $\Sigma$ can be written as
\be
ds^2_{\Sigma}=\gamma_{ab}dy^ady^b \, \, ,
\ee
where $\{y^{a}\}$ are the coordinates on $\Sigma$ (with $a=0, 1, 2$, while $\mu=0, 1, 2, 3$). The surface $\Sigma$ can be parametrized, both from the interior and from the exterior, as
\be
\Phi_{\pm}[x^{\mu}_{\pm}(y^a)]=0  \, \, .
\ee
The first set of matching conditions constitute the requirement that the three-dimensional induced metric $\gamma_{ab}$ must coincide when obtained from the two different sides of $\Sigma$. Since the induced metric is
\be
\gamma_{ab \pm}=\frac{\partial x^{\mu}_{\pm}}{\partial y^a} \frac{\partial x^{\nu}_{\pm}}{\partial y^b}g_{\mu \nu \pm} \, \, ,
\ee
in order for $\gamma_{ab \pm}$ to be the same on both sides, there must be a coordinates transformation on $\Sigma$ such that $\gamma_{ab \pm}=\gamma_{ab}$, or 
\be \label{firstjunction}
[\gamma_{ab}] \equiv \gamma_{ab +} - \gamma_{ab -}=0 \, \, ,
\ee
where the square brackets define the jump of a quantity across $\Sigma$. This condition guarantees that $g_{\mu \nu}$ is continuos everywhere on $\mathcal{M}$.

The second set of junction conditions concerns the first derivatives of $g_{\mu \nu}$ and their continuity. In order to obtain them, it useful to introduce the unit four-vector $n^{\mu}$ normal to $\Sigma$
\be
n_{\mu}=\frac{\partial \Phi/ \partial x^{\mu}}{\sqrt{g^{\alpha \beta}\frac{\partial \phi}{\partial x^\alpha}\frac{\partial \phi}{\partial x^\beta}}} \, \, ,
\ee
and the extrinsic curvature
\be
\begin{aligned}
K_{ab \pm}=&\frac{\partial x^{\mu}_\pm}{\partial y^a}\frac{\partial x^{\nu}_\pm}{\partial y^b}\nabla_{\mu}n_{\nu}=\\
=&-n_{\sigma} \left( \frac{\partial^{2}x^{\sigma}_{\pm}}{\partial y^{a} \partial y^{b}} + \Gamma_{\mu \nu \pm}^{\sigma}\frac{\partial x^{\mu}_{\pm}}{\partial y^{a}} \frac{\partial x^{\nu}_{\pm}}{\partial y^{a}} \right) \, \, ,
\end{aligned}
\ee
which we have expressed in terms of coordinates and Christoffel symbols $\Gamma_{\mu \nu \pm}^{\sigma}$ of $g_{\mu \nu \pm}$. The extrinsic curvature can be thought of as the \textit{geometric velocity} of $\Sigma$. In general, $G_{\mu \nu \pm}$ contains second derivative of the metric. If the first derivative of the metric are discontinuous across $\Sigma$, then the second derivatives can be written as a Dirac $\delta$ on $\Sigma$. We can consider a coordinate system such that $\Sigma$ is parametrized by $x^{\mu=3} \equiv x=0$. In the latter, the global energy-momentum tensor can be written as
\be \label{Tdecomposition}
T_{\mu \nu}=T_{\mu \nu +} \Theta(x) + T_{\mu \nu -} \Theta(-x) +S_{\mu \nu}\delta(x) \, \, ,
\ee
where $\Theta(x)$ is the Heavy-Side function, for which $d\Theta/dx=\delta(x)$, and $S_{\mu \nu}$ is the surface energy-momentum tensor of $\Sigma$, characterised by the property $S_{33}=S_{a3}=0$. Replacing \eqref{Tdecomposition} into the field equations we obtain the so-called Lanczos equations
\be
[K_{ab}]=S_{ab}-\frac{1}{2}\gamma_{ab}S \, \, .
\ee
If $T_{\mu \nu}$ has a discontinuity only across $\Sigma$, then $S_{ab}=0$ and
\be \label{secondjunction}
[K_{ab}]=0 \, \, .
\ee
This condition guarantees that the first derivatives of $g_{\mu \nu}$ are continuous across $\Sigma$.

If $g_{\mu \nu}$ is an arbitrarily spherically symmetric dynamical spacetime in coordinates $\{ \mathtt{t}, \mathtt{r},\theta,\phi \}$
\be \label{mostgeneralmetric}
ds^2=-A^2d\mathtt{t}^2+B^2d\mathtt{r}^2 +D^2d\Omega^2
\ee
where $A=A(\mathtt{t},\mathtt{r})=e^{\nu(\mathtt{t},\mathtt{r})}$, $B=B(\mathtt{t},\mathtt{r})=e^{\psi(\mathtt{t},\mathtt{r})}$, $D=D(\mathtt{t},\mathtt{r})$, then a spherical time-like hypersurface $\Sigma$ can be parametrized by the first spatial coordinate
\be \label{generalsigmaparametrization}
\Phi(\mathtt{t}, \mathtt{r}, \theta, \phi)=\mathtt{r}-\mathtt{R}_{b}(\mathtt{t})=0 \, \, .
\ee
The new font remarks that, in general, the time and radial coordinates are neither the comoving nor Schwarzschild ones. Moreover, the induced three-dimensional metric can be expressed using the coordinates $\{ \tau, \theta, \phi \}$
\be \label{inducedmetricpropertime}
ds^2_{\Sigma}=-d\tau^2+D_{b}(\tau)^2d\Omega^2 \, \, ,
\ee
where $\tau$ is the proper time on $\Sigma$, and $D_{b}(\mathtt{t})\equiv D[\mathtt{t}, \mathtt{R}_{b}(\mathtt{t})]$ for a generic time coordinate. Since \eqref{inducedmetricpropertime} and $ds^2_{\Sigma}$ computed from \eqref{mostgeneralmetric} must coincide, one can determine implicitly the relation between $\mathtt{t}$ and $\tau$, and that between $D$ and $D_{b}$.

Using $\tau$ as affine parameter to the describe the motion of $\Sigma$ itself, and making a suitable choice for $A_{\pm}$, $B_{\pm}$, $D_{\pm}$, after long calculations, one can determine $\gamma_{\tau \tau \pm}$, $\gamma_{\theta\theta \pm}$, $K_{\tau \tau \pm}$, and $K_{\theta\theta \pm}$ (indeed, in spherical symmetry only the $\tau\tau$ and $\theta\theta$ components are independent). In particular, using the comoving coordinates and the spacetime \eqref{FLRW} for $\mathcal{M}_{-}$, making the choice
\be
r-r_b=0
\ee
for \eqref{generalsigmaparametrization}, such that the two-dimentional reduction of $\Sigma$ coincides with the spatial region of spacetime permeated by the actual surface of the matter ball, and assuming a homogeneous equation of state for matter, we get
\begin{equation}
dt = d\tau \, \, , \quad
C(t, r_b) = D_b(\tau) \, \, ,
\end{equation}
\begin{equation}
\gamma_{\theta\theta -}= a(t)^2 r_{b}^2\, \, ,
\end{equation}
and
\begin{equation}
K_{\tau \tau}^-= 0 \, \, , \quad
K_{\theta\theta}^-= r_ba\sqrt{1-Kr_b^2} \, \, .
\end{equation}
Notably, \textit{if matter is homogeneous the comoving time coincides with the proper time on} $\Sigma$. Using the Schwarzschild coordinates and the spacetime \eqref{spacetimeexterior}-\eqref{lapseexterior} for $\mathcal{M}_{+}$,  making the choice
\be
R-R_b(T)=0
\ee
for \eqref{generalsigmaparametrization}, we get
\be \label{Schwarztime_propertimeonsigma}
dT = \left[f(R_b)-f(R_b)^{-1}\left(\frac{dR_b}{dT}\right)^2\right]^{-1/2} d\tau \, \, ,
\ee
\be
R_{b}[T(\tau)]= D_{b}(\tau) \, \, , \quad \gamma_{\theta\theta +}=R_b(T)^2 \, \, ,
\ee
and
\begin{equation}
\begin{aligned}
K_{\tau \tau + }&= -\frac{1}{2}\frac{2\ddot{R}_b+f_{,R}(R_b)}{\Delta(R_b)} \, \, , \\
K_{\theta\theta+}&= R_b\Delta(R_b) \, \, ,
\end{aligned}
\end{equation}  
where $\Delta(R_b) \equiv \sqrt{1-2M(R_b)/R_b+\dot{R}_b^2}$, and the dot indicates a derivative with respect to $\tau$. Obviously, $\gamma_{\tau \tau \pm}=-1$ always. Imposing the condition \eqref{firstjunction} for $\gamma_{\theta\theta}$, we obtain
\be \label{Schwarzbound_comovingbound}
R_{b}[T(t)]=r_{b}a(t) \, \, .
\ee
Using the latter, $t=\tau$, and \eqref{scalefactor} into \eqref{Schwarztime_propertimeonsigma}, we obtain
\be
T(t)=\int \frac{\sqrt{f(R_b)-V(a)r_b^2-Kr_b^2}}{f(R_{b})}dt \, \, .
\ee
As we expect from the dilation of the time intervals, for an arbitrarily far away observer with respect to the comoving observer, $T(t)$ is a monotonically increasing function. Finally, imposing the condition \eqref{secondjunction} for $K_{\theta\theta}$ (indeed, $[K_{\tau \tau}]=0$ does not give an independent equation), and using \eqref{Schwarzbound_comovingbound}, we obtain
\be
M(R_{b})= - \frac{r_{b}^2 R_{b}}{2} V(R_b) \, \, .
\ee
\\
In the OSD model, due to the Birkhoff-Jebsen uniqueness theorem, the exterior metric must be the Schwarzschild solution, which indeed satisfies the junction conditions with $M(R)=M_{0}$.

\end{document}